\documentclass[]{elsart}
\usepackage{psfig,latexsym}
\usepackage{amsmath}
\usepackage[sort&compress,numbers]{natbib}
\begin{document}
\journal{Physica A}
\date{6 September 2001}
\begin{frontmatter}
\title{Coordination of Decisions \\ in a Spatial Agent Model}
\author[rwc,hub]{Frank Schweitzer\thanksref{fs}} \thanks[fs]{Corresponding author,
  e-mail: schweitzer@gmd.de, http://ais.gmd.de/$^{\sim}$frank}
\author[rwc,ub]{J\"org Zimmermann}
\author[rwc]{Heinz M\"uhlenbein}
\address[rwc]{Real World Computing Partnership - Theoretical Foundation GMD
  Laboratory, Schloss Birlinghoven, D-53754 Sankt Augustin, Germany}
\address[hub]{Institute of Physics, Humboldt University,
  Invalidenstr. 110, 10115 Berlin, Germany}
\address[ub]{Institute of Computer Science III, University of Bonn,
R\"omerstr. 164, 53117 Bonn, Germany}

\begin{abstract}
  For a binary choice problem, the spatial coordination of decisions in
  an agent community is investigated both analytically and by means of
  stochastic computer simulations.  The individual decisions are based on
  different local information generated by the agents with a finite
  lifetime and disseminated in the system with a finite velocity.  We
  derive critical parameters for the emergence of minorities and
  majorities of agents making opposite decisions and investigate their
  spatial organization.  We find that dependent on two essential
  parameters describing the local impact and the spatial dissemination of
  information, either a definite stable minority/majority relation
  (single-attractor regime) or a broad range of possible values
  (multi-attractor regime) occurs. In the latter case, the outcome of the
  decision process becomes rather diverse and hard to predict, both with
  respect to the share of the majority and their spatial distribution.
  We further investigate how a dissemination of information on
  different time scales affects the outcome of the decision process. We
  find that a more ``efficient'' information exchange within a
  subpopulation provides a suitable way to stabilize their majority
  status and to reduce ``diversity'' and uncertainty in the decision
  process.
\end{abstract}
\end{frontmatter}
\newcommand{\mean}[1]{\left\langle #1 \right\rangle}
\newcommand{\abs}[1]{\left| #1 \right|}
\newcommand{\la}{\langle}
\newcommand{\ra}{\rangle}
\newcommand{\RA}{\Rightarrow}
\newcommand{\tet}{\vartheta}
\newcommand{\eps}{\varepsilon}
\newcommand{\bbox}[1]{\mbox{\boldmath $#1$}}
\newcommand{\ul}[1]{\underline{#1}}
\newcommand{\ol}[1]{\overline{#1}}
\newcommand{\non}{\nonumber \\}
\newcommand{\no}{\nonumber}
\newcommand{\eqn}[1]{eq. (\ref{#1})}
\newcommand{\Eqn}[1]{Eq. (\ref{#1})}
\newcommand{\eqs}[2]{eqs. (\ref{#1}), (\ref{#2})}
\newcommand{\pics}[2]{Figs. \ref{#1}, \ref{#2}}
\newcommand{\pic}[1]{Fig. \ref{#1}}
\newcommand{\sect}[1]{Sect. \ref{#1}}
\newcommand{\name}[1]{{\rm #1}}
\newcommand{\bib}[4]{\bibitem{#1} {\sc #2} (#4): #3.}
\newcommand{\vol}[1]{{\bf #1}}
\newcommand{\et}{{\it et al.}}
\newcommand{\einschub}[1]{\begin{center}%
           \parbox{15cm}{\sf\small #1}\end{center}}   
\newcommand{\fn}[1]{\footnote{ #1}}
\newcommand{\D}{\displaystyle}
\newcommand{\T}{\textstyle}
\newcommand{\SC}{\scriptstyle}
\newcommand{\SSC}{\scriptscriptstyle}
\renewcommand{\textfraction}{0.05}
\renewcommand{\topfraction}{0.95}
\renewcommand{\bottomfraction}{0.95}
\renewcommand{\floatpagefraction}{0.95}

\section{Introduction}
\label{1}

Decision making is one of the fundamental processes in economy but also
in social systems. If these systems consist of many interacting elements
-- which we call \emph{agents} here -- the system dynamics may be
described on two different levels: the \emph{microscopic} level, where
the decisions of the individual agents occur and the \emph{macroscopic}
level where a certain collective behavior can be observed. To find a link
between these two levels remains one of the challenges of complex systems
theory not only with respect to socio-economic systems \citep{haken-78,
  weidl-91, fs-ed-97, fs-silverberg-98-ed, arthur-et-97}.

Among the various factors that may influence the decision of agents we
mention the \emph{information} available on a particular subject -- such
as the price or the quality of a particular product, in an economic
context, or the benefits and harms that might result from the decision, in
a social context, but also information about the decisions of others.  A
somehow extreme example is given by the \emph{rational agent} model, one
of the standard paradigms of neoclassical economic theory. It is based on
the assumption of the agent's \emph{complete knowledge} of all possible
actions and their outcomes or a known probability distribution over
outcomes, and the \emph{common knowledge assumption}, i.e. that the agent
knows that all other agents know exactly what he/she knows and are
equally rational \citep{silverberg-verspagen-94}.

This implecitely requires an infinitely fast, loss-free and error-free
dissemination of information in the whole system. A more realistic
assumption would be based on the \emph{bounded rationality} of agents,
where decisions are not taken upon complete a priori information, but on
incomplete, limited knowledge 
distributed with finite velocity.  This however would require to model
the information flow between the agents explicitely.  A possibile
approach to this problem is given by the {\em spatio-temporal
  communication field} \cite{fs-holyst-00,fs-zimmermann-01}, that is also
used in this paper (cf. \sect{2}).  

Based on incomplete information, how does an agent make her decision on a
particular subject? The ``rational'' agent usually calculates her private
utility and tries to maximize it. Besides the methodological
complications involved e.g.  in the definitions of the utility functions
(that sometimes may anticipate the results observed in computer
simulations), it turns out in a world of uncertainty that the
maximization of private utilities can be only achieved by some
supplemented strategies.

In order to reduce the risk of making the wrong decision, it often seems
to be appropriate just to copy the decisions of others. Such an
\emph{imitation} strategy is widely found in biology, but also in
cultural evolution. Different species including humans imitate the
behavior of others of their species to become successful or just to adapt
to an existing community \citep{dugatkin-01}.  A physically visible
example of such an imitation strategy is the formation of trails commonly
created among humans, hoofed animals or social insects. Very similar to
the model discussed in the present paper, this can be simulated in an
agent model based on a non-linear positive feedback between the agents,
where sometimes different kind of information is involved
\citep{fs-lao-family-97,helbing-fs-et-97}.

But imitation strategies are also most powerful in economic systems,
where late entrants quite often size markets from pioneers
\citep{schnaars-94}. While the latter ones are forced to spend heavily on
both product and market development, imitators can often profit from an
already existing market, while avoiding the risk of making costly
mistakes.  In the case, where agents can observe the \emph{payoffs}
generated by other agents, \emph{information contagion}
\citep{arthur-lane-93} has been presented as an explanation for
particular patterns of macrobehavior in economic systems, for example
path-dependence and lock-in-effects \citep{lane-vescovini-96, vriend-95}.
This has been also simulated within the concept of \emph{social
  percolation} \citep{solomon-et-00, weisbuch-stauffer-00}, which may
explain the occurrence of extreme market shares, based on positive
feedback processes between adaptive economic agents.

Information contagion however involves the transmission of two different
information, the \emph{decision} made by an agent and the \emph{payoff}
received. The situation becomes different when agents only observe the
choices of other agents and tend to imitate them, \emph{without} complete
information about the possible consequences of their choices. This is
commonly denoted as \emph{herding behavior} which plays a considerable
role in economic systems \citep{banaerjee-92, kirman-93}, in particular
in financial markets \citep{lux-95}, but also in human and biological
systems where \emph{panic} can be observed \citep{helbing-panic-00}.

In social systems, herding behavior may result from the many (internal or
external) interdepencencies of an agent community that push or pull the
individual decision into a certain direction, such as peer pressure or
external influences. The \emph{social impact theory} \citep{latane-81,
  nowak-szam-latane-90} that intends to describe the transition from
``private attitude to public opinion'' has covered some these collective
effects in a way that can be also formalized within a physical approach
\citep{kohring96, lewenst-nowak-latane-92, holyst-kacp-fs-01}.

One of the modelling impacts of the social impact theory was the
``rediscovering of physical space'' in sociology, i.e. distance matters
for social influence \citep{hillier-hanson-90, latane-et-95,
  nowak-latane-lewenstein-94}.  Hence, instead of mean-field approaches
where all actions of agents are coupled via a mean field, spatial models
become of increasing interest. In addition to the question of how
individual decisions of agents may affect the macrobehavior of the
system, now the question becomes important how these desisions may
organize themselves \emph{in space}, i.e. what kind of \emph{spatial
  patterns} may be observed on the global level.

The most commonly ``spatial'' model used is based on the cellular
automaton approach, i.e. each agent occupies a lattice site. Often the
lattice is also fully covered by agents, i.e. there is a homogeneous
distribution in ``space''. As a third assumtion mostly used, only
nearest-neighbor interactions are included, however the social impact
theory has also provided an ansatz, where the interaction between each
two agents can be considered in a weighted manner (different from the
mean field). 

Within the framework of cellular automata different models of collective
opinion formation and group decision processes have been introduced, such
as ising-like models \citep{galam-97, galam-zucker-00, sznajd-00} or
models based on a modified social impact theory \citep{kacp-holyst-96,
  kacp-holyst-97, holyst-kacp-fs-00} that have been also applied to the
coordination of individual economic decisions \citep{nowak-et-00}.  Also
sociological problems such as the formation of support networks
\citep{hegselmann-flache-98} have been simulated within an cellular
automaton approach.  The role of local information exchange has been
further considered within the framework of the Minority Game
\citep{challet-zhang-98}, where the emergence of rich and poor spatial
domaines could be shown \citep{slanina-00}. Here, however, only
nearest-neighbor interactions have been taken into accout (more
precisely, only interactions with the left-hand neighbor in a linear
chain model).

Taking into account the different perspectives on decision processes
outlined above, we may conclude the following requirements for a modeling
approach: It should be \textbf{(i)} an \emph{agent-based} approach that
allows to simulate individual decisions, \textbf{(ii)} a \emph{spatial}
model that takes into account physical distances between agents, but
\textbf{(iii)} is \emph{not} restricted to nearest-neighbor or mean-field
interaction.  Further it should allow \textbf{(iv)} a
\emph{heterogeneous} spatial distribution of both agents and information
and \textbf{(v)} an explicite modeling of \emph{information exchange}.
Regarding the decision process, the model should take into accout
\textbf{(vi)} the influence of information \emph{locally} available to
the agent (instead of a common knowledge assumption), \textbf{(vii)}
strategic elements such as \emph{imitation} that go beyond the
calculation of a private utility.

In the next section, we want to propose an agent model, that is novel in
the sense that it meets all of the requirements listed above -- at least
to some extent -- and is further simple enough to serve as a toy model
for investigating spatial effects in decision processes both analytically
and by means of computer simulations.

As we will show, this model describes the emergence of a majority and a
minority of agents making the same decision (\sect{3}). But besides the
existence of a global majority, there are regions that are dominated by
the minority, hence a \emph{spatial coordination} of decisions among the
agents occurs (\sect{4}).

The main result of this paper, however, is \emph{not} just the
observation of the clustering of opinions, which is a pervasive feature
of a wide range of models -- although in economics this phenomenon is
very often described only on the aggregated (macro) level, but not on the
(micro) level of the agents. In this paper, we derive conditions
(\sect{4}) under which -- for the same given parameters -- a large
variety of possible spatial decision patterns can be found, and the
outcome of the decision process becomes certainly unpredictable,
\emph{both} with respect to the share of the majority \emph{and} to the
spatial distribution.  We further show how the exchange of information
affects the possibility of a broad range of spatial decision patterns
(\sect{5}). In particular, we discuss how ``efficient'' information
exchange provides a suitable way to stabilize the majority status of a
particular subpopulation -- or to avoid ``diversity'' and uncertainty in
the decision process (\sect{5}).

\section{Toy Model of Communicating Agents}
\label{2}

Let us consider a 2-dimensional spatial system with the total area $A$,
where a
community of $N$ individuals 
exists.  In general, $N$ can be changed by birth and death processes but
$A$ is assumed fixed. In a more abstract sense, each individual $i$ shall
be treated as an \emph{agent}, i.e. a rather autonomous entity which is
assigned two individual parameters: its position in space, $r_{i}$, which
should be a continuous variable, and its current ``opinion'',
$\theta_{i}$ (with respect to a definite aspect or problem). The latter
one is a discrete valued parameter representing an \emph{internal degree
  of freedom} (which is a rather general view of ``opinion'').

To be more specific, let us discuss for instance the separate disposal of
recyclable material.  Each agent in the system needs to decide whether she
will cooperate in the recycling campaign or defect. Then, there are only
two (opposite) opinions, i.e.  $\theta_{i} \in \{+1,-1\}$, or $\{+,-\}$
to be short.  $\{+\}$ shall indicate the cooperating agent, and $\{-\}$
the defecting agent.

From the classical economic perspective, the agents' decision about her
opinion may depend on an estimate of her \emph{utility}, i.e. what she
may gain compared to her own effort, if she decides to cooperate or not.
Here, we neglect any question of utility and may simply assume that the
agent will more likely do what others do with respect to the specific
problem, i.e. she will decide to cooperate in the recycling campaign if
most of her neighbors will do so, and defect if most of their neighbors
have the same opinion in this case. This type of herding behavior in
decision processes -- a special kind of the imitation strategy -- is well
known from different fields, as discussed in \sect{1}.

This example raises the question about the interaction between agents at
different locations, i.e. how is agent $i$ at position $r_{i}$ affected
by the decisions of other agents at closer or far distant locations?  In
a checkerboard world, commonly denoted as cellular automaton, a common
assumption is to consider only the influence of agents, which are at the
(four or eight) nearest neigbour sites or also at the second-nearest
neighbor sites, etc. Contrary, in a \emph{mean-field approximation}, all
agents are considered as influencial via a mean field, which affects each
agent at the same time in the same manner.

Our approach will be different from these ones in that we will consider a
continuous space and a gradual, time delayed interaction between
all agents. We assume that agent $i$ at position $r_{i}$ is not directly
affected by the decisions of other agents, but only receives information
about their decisions via a \emph{communication field} generated by the
agents with the different opinions. This field is assumed a
scalar {\em multi-component spatio-temporal field}
$h_{\theta}(\bbox{r},t)$, which obeys the following equation:
\begin{equation} 
\label{hrt} 
\frac{\partial}{\partial t} h_{\theta}(\bbox{r},t) = 
\sum_{i=1}^{N}s_{i}\;\delta_{\theta,\theta_{i}}\;
\delta(\bbox{r}-\bbox{r}_{i})\;
- \;k_{\theta} h_{\theta}(\bbox{r},t) \;+ \;D_{\theta} 
\Delta h_{\theta}(\bbox{r},t). 
\end{equation} 
Every agent contributes permanently to this field with her personal
``strength'' or influence, $s_{i}$.  Here, $\delta_{\theta,\theta_{i}}$
is the \name{Kronecker} Delta indicating that the agents contribute only
to the field component which matches their opinion $\theta_{i}$.
$\delta(\bbox{r}-\bbox{r}_{i})$ means \name{Dirac's} Delta function used
for continuous variables, which indicates that the agents contribute to
the field only at their current position, $\bbox{r}_{i}$.

The \emph{information} generated this way has a certain life time
$1/k_{\theta}$, further it can spread throughout the system in a diffusion-like
process, where $D_{\theta}$ represents the diffusion constant
for information exchange.  We have to take into account that there are
two different opinions in the system, hence the communication field
should also consist of two components, $\theta=\{+1,-1\}$, each
representing one opinion. Note, that the parameters describing the
communication field, $s_{i}$, $k_{\theta}$, $D_{\theta}$ do not
necessarily have to be the same for the two opinions.

The {\em spatio-temporal communication field} $h_{\theta}(\bbox{r},t)$ is
used to reflect some important features of communication in social
systems:
\begin{itemize}
\item[(i)] the existence of a {\em memory}, which reflects the past
  experience. In our model, this memory exist as an external memory, the
  lifetime of which is determined by the decay rate of the field,
  $k_{\theta}$.
\item[(ii)] an {\em exchange of information} in the community with a {\em
    finite} velocity. It means that the information will eventually reach
  each agent in the whole system, but of course at different times.
\item[(iii)] the influence of {\em spatial distances} between agents.
  Thus, the information generated by a specific agent at position $r_{i}$
  will affect agents at a closer spatial distance earlier and thus with
  larger weight, compared to far distant agents.
\end{itemize}

The communication field $h_{\theta}(\bbox{r},t)$ influences the agent's
decisions as follows: At a certain location $\bbox{r}_{i}$ agent $i$ with
e.g. opinion $\theta_{i}=+1$ is affected by two kinds of information: the
information $h_{\theta=+1}(\bbox{r}_{i},t)$ resulting from agents who
share her opinion, and the information $h_{\theta=-1}(\bbox{r}_{i},t)$
resulting from the opponents. The diffusion constants $D_{\theta}$
determine how fast she will receive any information, and the decay rate
$k_{\theta}$ determines, how long a generated information will exist.
Dependent on the information received \emph{locally}, the agent has two
opportunities to act: she can \emph{change her opinion} or she can keep
it. A possible ansatz for the transition rate to change the opinion reads
\citep{fs-holyst-00}:
\begin{equation} 
w(-\theta_{i}|\theta_{i})=  
\eta \, \exp\left\{ -
  \frac{h_{\theta}(\bbox{r}_{i},t)-h_{-\theta}(\bbox{r}_{i},t)}{T}\right\}\;;  \quad
w(\theta_{i}|\theta_{i}) = 0 
\label{wh} 
\end{equation} 
The probability to change opinion $\theta_{i}$ is rather small, if the
local field $h_{\theta}(\bbox{r}_{i},t)$, which is related to the support
of opinion $\theta_{i}$, overcomes the local influence of the opposite
opinion.  Here, $\eta$ defines the time scale of the transitions.  $T$ is
a parameter which represents the \emph{erratic circumstances} of the
opinion change, based on an incomplete or incorrect transmission of
information. Note, that $T$ is measured in units of the communication
field.  In the limit \mbox{$T \to 0$} the opinion change rests only on
the difference $\Delta h (\bbox{r}_{i},t)=
h_{\theta}(\bbox{r}_{i},t)-h_{\theta'}(\bbox{r}_{i},t)$, leading to
``rational'' decisions (cf. also \citep{galam-97}), i.e. decisions that
are totally determined by the external information. In the limit \mbox{$T
  \to \infty$}, on the other hand, the influence of the information
received is attenuated, leading to ``random'' decisions.  We note that
$T$ can be also interpreted in terms of a ``social temperature''
\citep{kacp-holyst-96,kacp-holyst-97}, i.e.  it is a measure for the
randomness in social interaction.
  
For $N= \mathrm{const.}$, the community of agents may be described by the
time-dependent canonical $N$-particle distribution function
\begin{equation} 
\label{prt} 
P(\ul{\theta},\ul{r},t)=P(\theta_{1},\bbox{r}_{1},...,\theta_{N},\bbox{r}_{N},t), 
\end{equation} 
which gives the probability to find the $N$ agents with the opinions
$\theta_{1},...,\theta_{N}$ in the vicinity of
$\bbox{r}_1,....,\bbox{r}_N$ on the surface $A$ at time $t$. The time
depentent change of $P(\ul{\theta},\ul{r},t)$ is then given by the
following master equation \citep{weidl-91}:
\begin{equation} 
\frac{\D \partial}{\D \partial t}P(\ul{\theta},\ul{r},t) =   
\sum_{\ul{\theta'} \neq \ul{\theta}} \Big[
w(\ul{\theta}|\ul{\theta}') P(\ul{\theta}',\ul{r},t) -  
w(\ul{\theta}'|\ul{\theta}) P(\ul{\theta},\ul{r},t) \Big] 
\label{master} 
\end{equation} 
\Eqn{master} describes the ``gain'' and ``loss'' of agents with the
coordinates $\bbox{r}_1,...,\bbox{r}_N$ due to opinion changes, where
$w(\ul{\theta}|\ul{\theta}')$ means any possible transition within the
opinion distribution $\ul{\theta}'$ which leads to the assumed
distribution $\ul{\theta}$.  \Eqn{master} together with \eqs{hrt}{wh}
forms a complete description of our system, which depends on the
parameters describing the agent density, i.e. $N$, $A$, and the
components of the communication field, $s_{i}$, $k_{\theta}$,
$D_{\theta}$. In order to find possible solutions of the master equation,
we will use computer simulations, and in particular apply the stochastic
simulation technique \citep{eb-feistel-82,fs-98-jcs}.  But before
investigating the spatially distributed system, we first will discuss a
mean-field approximation, in order to get some insight into the complex
dynamics of the agent system.


\section{Mean-Field Approach}
\label{3}
In this section, we will neglect any spatial effects of the agents
distribution and the communication field.  This case, which has been
discussed in more detail also in \citep{fs-holyst-00}, may have some
practical relevance for communities existing in small systems with short
distances between different agents. In particular, in such small
communities a very fast exchange of information may hold, i.e.  spatial
inhomogenities in the communication field are equalized immediately.
Thus, in this section, the discussion can be restricted to subpopulations
with a certain opinion rather than to agents at particular locations.

Let us define the share $x_{\theta}$ of a subpopulation $\theta$ and the
respective mean density $\bar{n}_{\theta}$ in a system of size $A$
consisting of $N$ agents:
\begin{equation}
  \label{fraction}
  x_{\theta}(t)=\frac{N_{\theta}(t)}{N}\;;\quad 
\bar{n}_{\theta}(t) = \frac{N_{\theta}(t)}{A}
\end{equation}
where the total number of agents sharing opinion $\theta$ at time $t$
fulfils the condition
\begin{equation}
\label{sum}
\sum\nolimits_{\theta}N_{\theta}(t)=N_{+}(t)+N_{-}(t)=N={\rm const.} 
\;;\quad x_{+}(t)=1-x_{-}(t)    
\end{equation} 
In the mean-field approach, the communication field
$h_{\theta}(\bbox{r},t)$ can be approximated by a mean value
$\bar{h}_{\theta}(t)$ which obeys the following dynamic equation:
\begin{equation} 
\frac{\partial \bar{h}_{\theta}(t)}{\partial t} = 
- k_{\theta}\bar{h}_{\theta}(t) + s_{\theta} \bar{n}_{\theta} 
\label{hat-t} 
\end{equation} 
Here, we have assumed that agents with the same opinion $\theta$ will
have the same influence $s_{i}\to s_{\theta}$. We note that the case of a
``strong leader'', where one agent has a personal strength $s_{l}$ much
larger than the usual strength $s_{i}$ has been discussed in
\citep{kacp-holyst-97, fs-holyst-00, kacp-holyst-00}.

The dynamic equation for the size of subpopulation $\theta$ can be
derived from \eqn{master} in the mean--field approximation as follows
\citep{fs-holyst-00}: 
\begin{eqnarray} 
\label{xdet} 
\dot{x}_{\theta} &=&(1-x_{\theta})\,\eta \, 
\exp(a) - x_{\theta}\,\eta \, \exp(-a)\;; \quad 
a= \left[\bar{h}_{\theta}(t)-\bar{h}_{-\theta}(t)\right]/ T
\end{eqnarray} 
Via $\Delta \bar{h}(t)= \bar{h}_{\theta}(t)-\bar{h}_{-\theta}(t)$, this
equation is coupled to \eqn{hat-t}.

Let us now investigate the possible existence of stationary states,
$\dot{x}_{\theta}=0$, $\dot{h}_{\theta}=0$. For the two field components we
find from \eqn{hat-t} with $\bar{n}=N/A$: 
\begin{equation} 
\label{ha} 
\bar{h}_{+}^{stat}=\frac{\D  s_{+}}{\D k_{+}}\,\bar{n}x_{+} \;; \quad  
\bar{h}_{-}^{stat}=\frac{\D s_{-}}{\D k_{-}}\,\bar{n}(1-x_{+})
\end{equation} 
Let us for the moment assume that the parameters of both field components
are identical, i.e. $s_{+}=s_{-}\equiv s$, $k_{+}=k_{-}\equiv k$, a
more complex case will be discussed in Sect. {5}. Then, the value $a$,
\eqn{xdet}, which depends on the difference between the field components,
can in the stationary limit be expressed as:
\begin{equation}
  \label{a-stat}
a=\kappa \left(x_{+}-\frac{\D 1}{\D 2}\right) \;\;\mbox{with} \;\;
\kappa=\frac{\D 2s\,\bar{n}}{\D k\, T} 
\end{equation}
The parameter $\kappa$ plays the role of a bifurcation
parameter that 
includes the specific \emph{internal conditions} within the community,
such as the population density, the individual strength of the opinions,
the life time of the information generated or the randomness $T$.
Combining \eqs{xdet}{a-stat},  the condition for
stationary solutions can be expressed by means of a function
$\mathcal{L}(x,\lambda)$ and $\kappa=\lambda/2$ as
\begin{equation}
  \label{xl-stat}
\mathcal{L}(x_{+},2\kappa)=0 
\end{equation}
 where $x \in (0,1)$ and 
$\mathcal{L}(x,\lambda)$ is defined as: 
\begin{equation}
  \label{l-xl-def}
\mathcal{L}(x,\lambda):=\ln{\frac{x}{1-x}} -
\lambda \left(x-\frac{1}{2}\right)  
\end{equation}

\begin{figure}[ht] 
\centerline{\psfig{figure=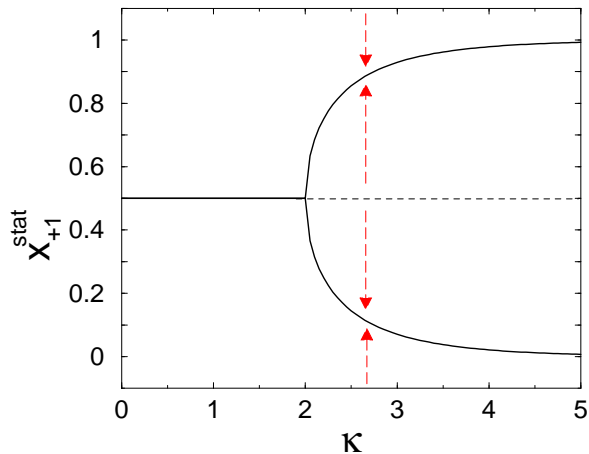,width=7.5cm}} 
\caption[fig1]{ 
 Stationary solutions for $x_{+}$ (eq. \ref{xl-stat}) for 
  different values of $\kappa$. The bifurcation at the critical value 
  $\kappa^{c}=2$ is clearly visible. For $\kappa=2.66$ that is used for
  some of the computer simulations, we find in the mean field limit the
  stationary values $x_{+}=0.885$ and $x_{+}=0.115$ for the majority and
  the minority status, respectively.  
  \label{crit}} 
\end{figure} 

In \citep{fs-holyst-00} we found that, depending on $\kappa$, different
stationary values for the fraction of the subpopulations exist (cf. also
\pic{crit}).  For $\kappa <2$, $x_{+}=0.5$ is the only stationary
solution, which means a stable community where both opposite opinions
have the same influence. However, for $\kappa>2$, the equal distribution
of opinions becomes unstable, and a separation process towards a
preferred opinion is obtained, where $x_{\pm}=0.5$ plays the role of a
separatrix. Then, two stable solutions are found where both opinions
coexist with different shares in the community, as shown in \pic{crit}.
Hence, each subpopulation can exist either as a \emph{majority} or as a
\emph{minority} within the community. Which of these two possible
situations is realized, depends in a deterministic approach on the
initial fraction of the subpopulation. For initial values of $x_{+}$
below the separatrix, $0.5$, the minority status will be most likely the
stable situation, as shown in \citep{fs-holyst-00}.

The share of the majority or minority for a given value of $\kappa>2$ can
be implicitely calculated from \eqn{xl-stat}. The
critical value for $\kappa$ where the bifurcation occurs is determined by \eqn{xl-stat} together with:
\begin{equation}
  \label{kappa}
\left. \frac{\partial \mathcal{L}(x_{+},2\kappa)}{\partial
  x_{+}}\right|_{x_{+}=1/2} = 4 - 2\kappa := 0
\end{equation}
which results in $\kappa^{c}=2$.  From this condition we can derive by
means of \eqn{a-stat} a \emph{critical population size},
\begin{equation} 
  \label{n-crit} 
 N^{c}= k\,A\,T/s,  
\end{equation} 
where for larger populations an equal fraction of opposite opinions is
certainly unstable. We note that this critical value has been derived
based on a mean field analysis and therefore does not consider finite
size or discrete effects.

If we consider e.g. a \emph{growing community} with fast communication,
then both contradicting opinions are balanced, as long as the population
number is small. However, for $N>N^{c}$, i.e.  after a certain population
growth, the community tends towards one of these opinions, thus
necessarily separating into a majority and a minority. Note that
\eqn{n-crit} for the critical population size can be also interpreted in
terms of a critical social temperature, which leads to an opinion
separation in the community. This has been discussed in more detail in
\citep{fs-holyst-00}.
 
From \pic{crit}, we see further, that the stable coexistence between
majority and minority breaks down at a certain value of $\kappa$, where
almost the whole community shares the same opinion. From \eqn{xl-stat} it
is easy to find that e.g.  $\kappa \approx 4.7$ yields $x_{\theta}\approx
\{0.01;0.99\}$, which means that about $99\%$ of the community share
either opinion $+1$ or $-1$. For smaller values of $\kappa$, for instance
for $\kappa=2.66$, we find $x_{+}=0.885$ for the majority status, and
$x_{+}=0.115$ for the minority status respectively. Of course, due to
the symmetry between both opinions, the opposite relation, $x_{+}=0.115$
and $x_{-}=0.885$ is possible with the same probability. Which one is
realized may depend on the fluctuation during the early stage of the
evolution of the agent system. \pic{w0-w1} shows a particular realization
obtained from computer simulation of 400 agents who at $t=0$ are
randomly assigned one of the opinions $\{+1,-1\}$. 

\begin{figure}[ht] 
\centerline{\psfig{figure=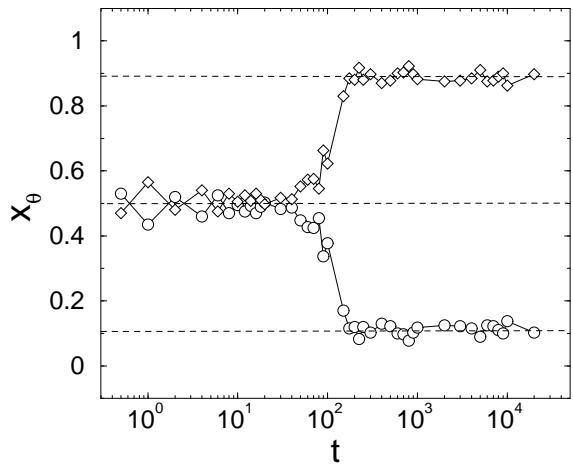,width=7.5cm}} 
\caption[fig2]{
  Computer simulation of the relative subpopulation sizes $x_{+}$
  ($\Diamond$)  and $x_{-}$ ($\circ$) vs. time $t$ for a community of
  $N=400$ agents. Parameters: $A=400$, $s=0.1$, $k=0.1$, $T=0.75$, i.e.
  $\kappa=2.66$. Initially, each agent has been randomly assigned opinion
  $+1$ or $-1$. The dashed lines indicate the inital equal distribution
  ($x_{\theta}=0.5$) and the minority and majority sizes
  ($x_{\theta}=\{0.115;0.885\}$) which follow from \eqn{xl-stat}.
\label{w0-w1}} 
\end{figure} 
\begin{figure}[ht] 
\centerline{\psfig{figure=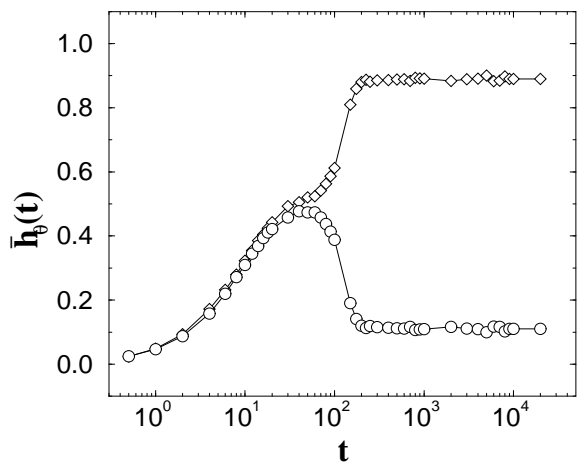,width=7.5cm}} 
\caption[fig1]{ 
  Components $h_{+}$ ($\Diamond$) and $h_{-}$ ($\circ$) of the mean
  communication field vs. time $t$ for the simulation shown in
  \pic{w0-w1}. After the initial time lag $t^{\star}\sim 5/k=50$, both
  field components evolve differently.
  \label{ht}} 
\end{figure} 

As indicated in \pic{w0-w1}, there is a latent period in the beginning
\emph{before} the minority/majority relation emerges, i.e. during this
period it is not clear which one of the two subpopulations, the
cooperators or the defectors will gain the majority status.  This initial
time lag $t^{\star}$ can be roughly estimated as $t^{\star}\sim 5/k=50$
\citep{fs-lsg-94}, where $k$ is the decay constant of the field.
$t^{\star}$ is needed to establish the communication field, as shown in
\pic{ht}. With the same set of parameters, both components of the
communication field at first evolve in the same manner.  However after
the initial time lag, a selection in favor of one of the components
occurs which then breaks the symmetry between both opinions.  Hence, the
communication field plays the role of an \emph{order parameter} as known
from synergetics \citep{haken-78}.  Consequently, for $t\geq t^{\star}$, a
transition from the unstable equal distribution between both opinions
toward a majority/minority relation is clearly visible in \pic{w0-w1}.
The time period to eventually establish this relation is then rather
short, since the case discussed in this section is related to a very fast
exchange of information.

Eventually, we want to point out that the symmetry between the two
opinions can be broken due to external influences on the agents. In
\citep{fs-holyst-00} we have considered two similar cases: (i) the
existence of a {\em strong leader} in the community, who possesses a
strength $s_{l}$ which is much larger than the usual strength $s$ of the
other individuals, (ii) the existence of an external field, which may
result from government policy, mass media, etc.  which support a certain
opinion with a strength $s_{m}$.  The additional influence
\mbox{$s_{ext}:=\{s_{l}/A,s_{m}/A\}$} mainly effects the mean
communication field, eq.  (\ref{hat-t}), due to an extra contribution,
normalized by the system size $A$. We found within the mean-field
approach that at a critical value of $s_{ext}$, the possibility of a
minority status completely vanishes. Hence, for a certain supercritical
external support, the supported subpopulation will grow towards a
majority, regardless of its initial population size, with no chance for
the opposite opinion to be established. This situation is quite often
realized in communities with one strong political or religious leader
(``fundamentalistic dictatorships''), or in communities driven by
external forces, such as financial or military power (``banana
republics'').

\section{Spatial Information Dissemination}
\label{4}
\subsection{Results of Computer Simulations}

\label{4.1}

The previons section has shown within a mean-field approach the emergence
of a minority/majority relation in the agents community.  With respect to
the example of  the recycling campaign adressed in the beginning, it means
that \emph{either} most of the agents decide to cooperate \emph{or}
most of them defect.

The question remains how the cooperators and the defectors organize
themselve in space by means of a spatial dissemination of information. In
order to consider the spatial dimension of the system explicitely, let us
start with the previous example of $N=400$ agents randomly distributed in
a system of size $A$ (cf. \pic{snap}), with random initial opinions. They
get information about the opinions of other agents by means of the
two-component communication field $h_{\theta}(\bbox{r},t)$, \eqn{hrt},
which now explicitely considers space and therefore ``diffusion'' of
information. The two-dimensional system is here treated as a torus, i.e.
we assume periodic boundary conditions.

In this section, we assume again that the parameters decribing the
communication field, are the same for both components, i.e.
\begin{equation}
  \label{equal}
s_{+}=s_{-}\equiv s\;;\;\;
k_{+}=k_{-}\equiv k\;;\;\;
D_{+}=D_{-}\equiv D  
\end{equation}
a different case will be investigated in \sect{5}.  \pic{snap} shows
three snapshots of the spatial distribution of the cooperators and the
defectors, while \pic{w0-w1-06} shows the respective evolution of the
subpopulation shares. Evidently, we find again the emergence of a
majority/minority relation -- this time however, on a larger time scale
compared to \pic{w0-w1}, which is basically detemined by the information
diffusion, expressed in terms of $D$.
But the initial latent time lag for the emergence of the
majority/minority relation is about the same, which is needed again to
establish the communication field.

\begin{figure}[ht] 
\centerline{
\psfig{figure=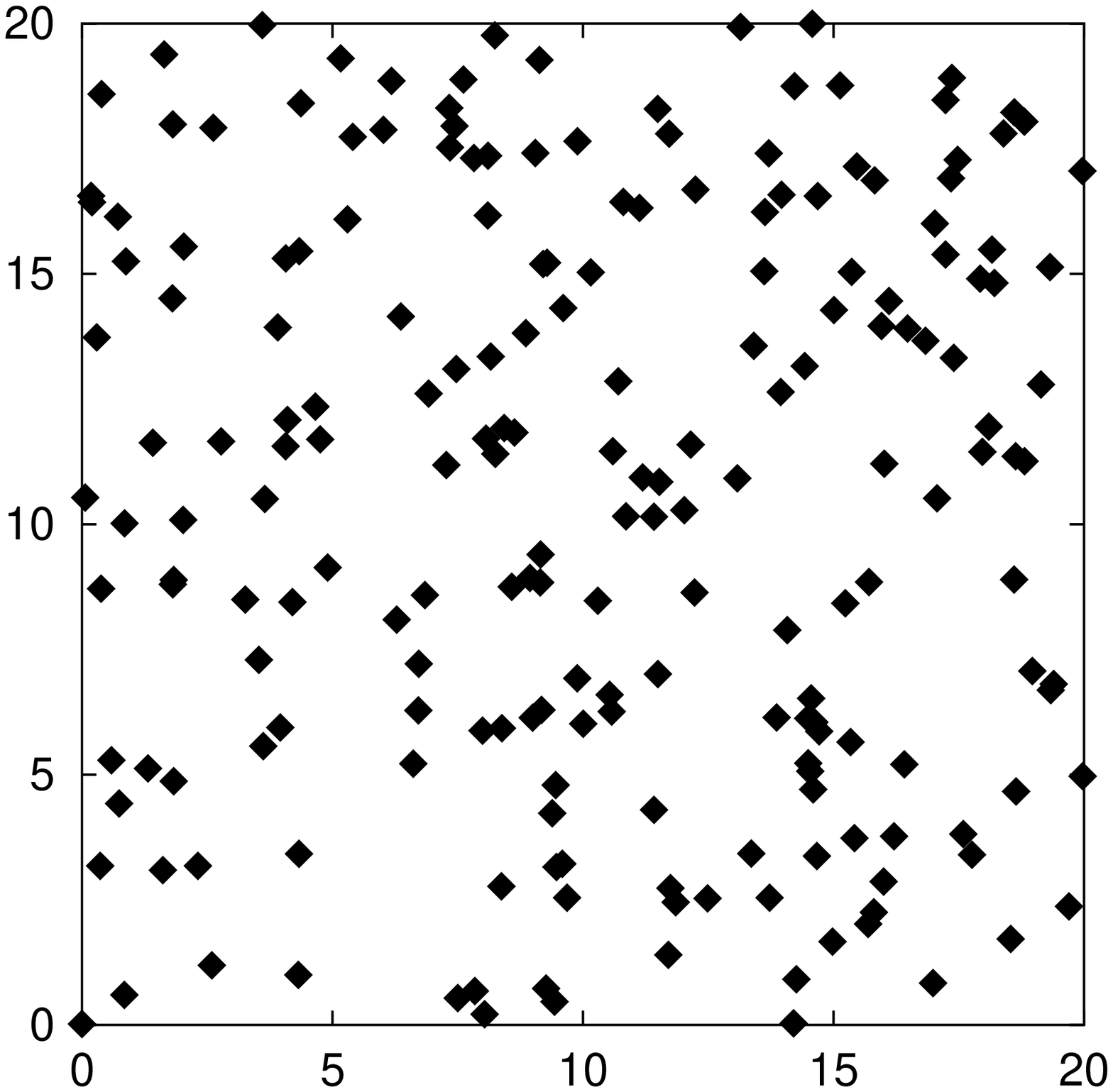,width=5.0cm,angle=-90}
\hspace{0.4cm} (a) 
\hspace{0.4cm}
\psfig{figure=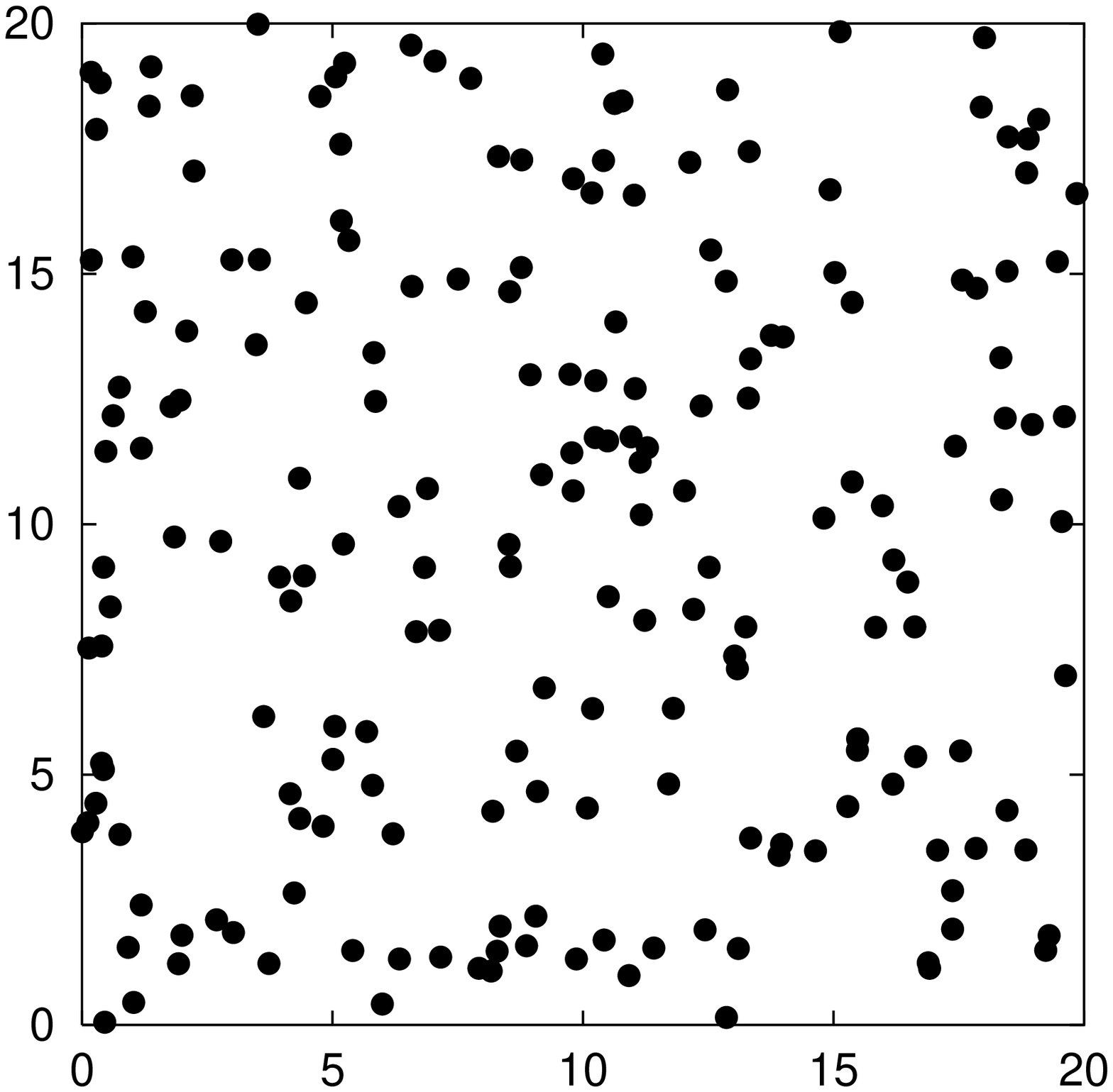,width=5.0cm,angle=-90}}
\vspace{.8cm}
\centerline{
\psfig{figure=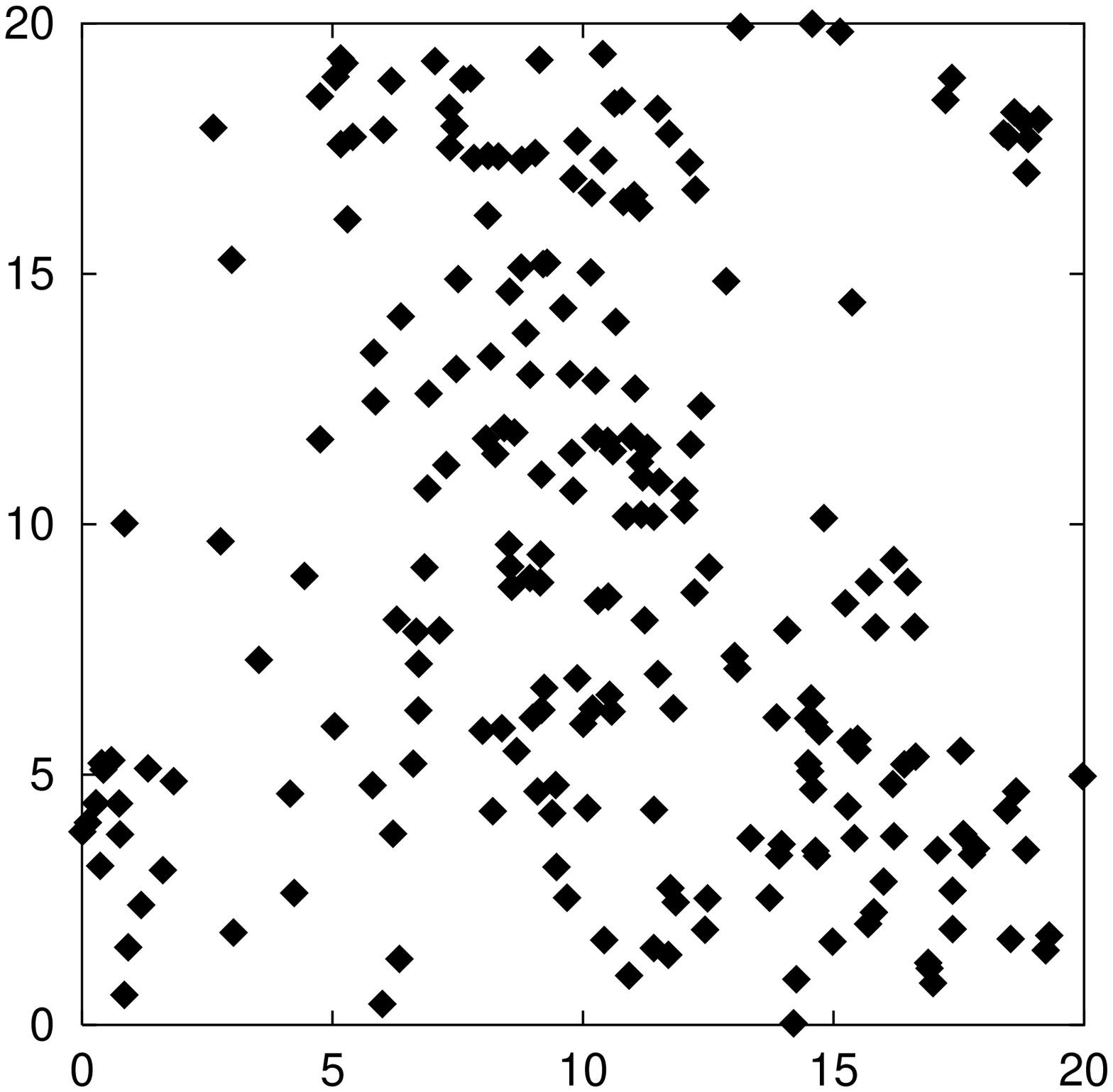,width=5.0cm,angle=-90}
\hspace{0.4cm} (b) 
\hspace{0.4cm}
\psfig{figure=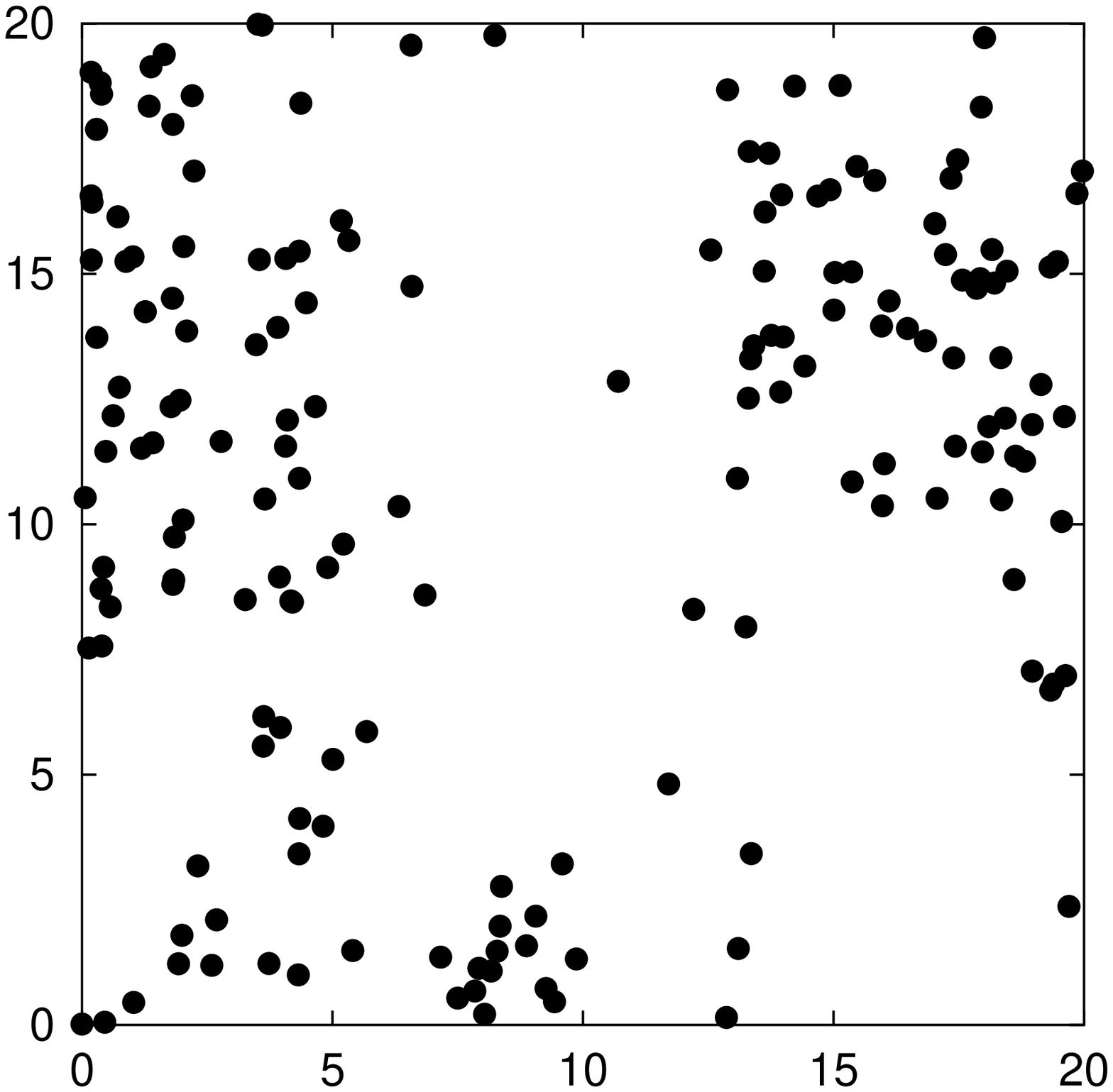,width=5.0cm,angle=-90}}
\vspace{.8cm}
\centerline{
\psfig{figure=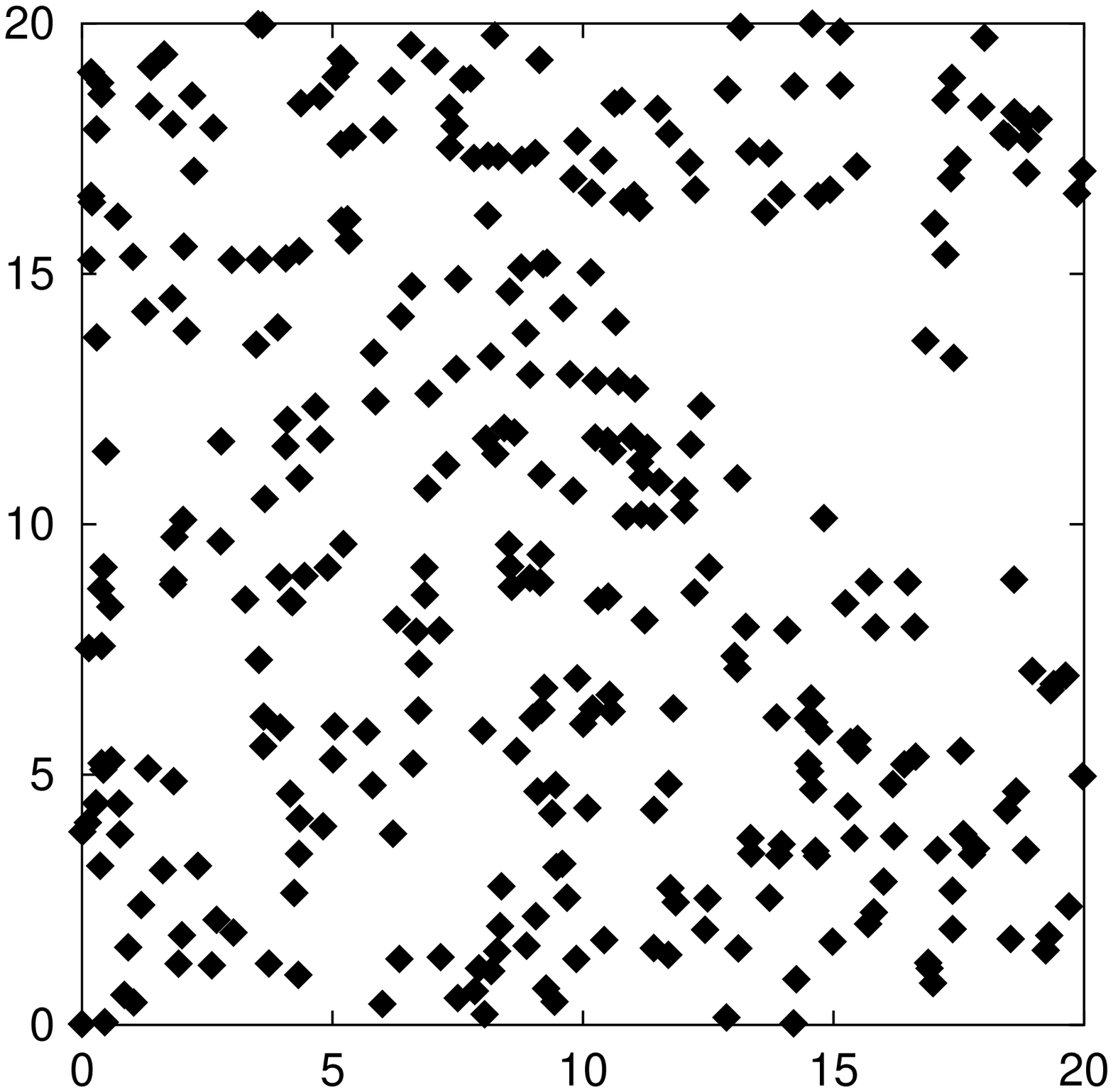,width=5.0cm,angle=-90}
\hspace{0.4cm} (c) 
\hspace{0.4cm}
\psfig{figure=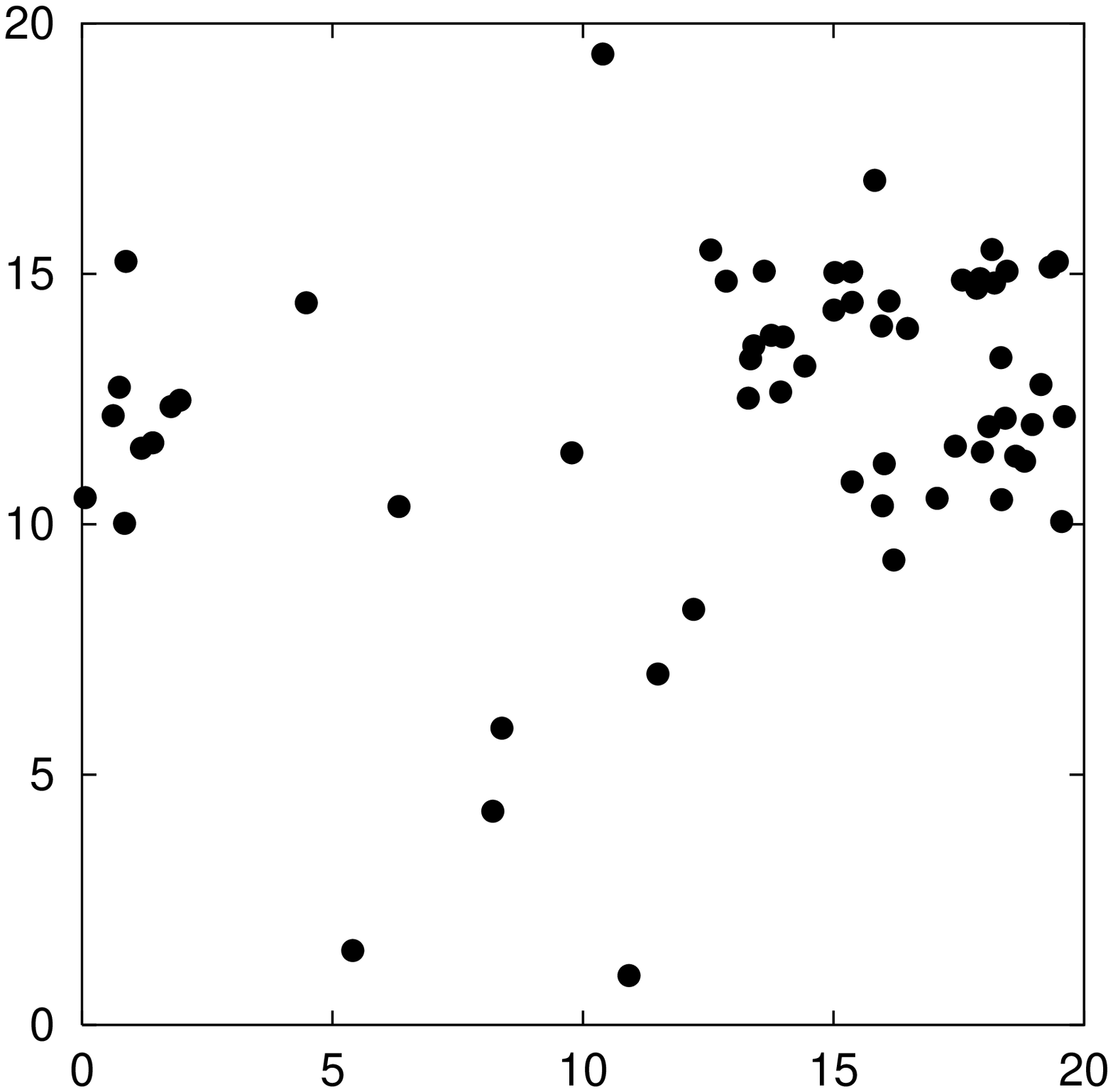,width=5.0cm,angle=-90}}
\caption[fig3]{
Computer simulations of the spatial distribution of cooperators
  ($\Diamond$ left column) and defectors ($\circ$ right column).  The
  snapshots are taken at three different times: (a) $t=10^{0}$, (b)
  $t=10^{2}$, (c) $t=10^{4}$. For the parameters and initial conditions
  see \pic{w0-w1}, additionally $D=0.06$.
\label{snap}} 
\end{figure} 
\begin{figure}[ht] 
  \centerline{\psfig{figure=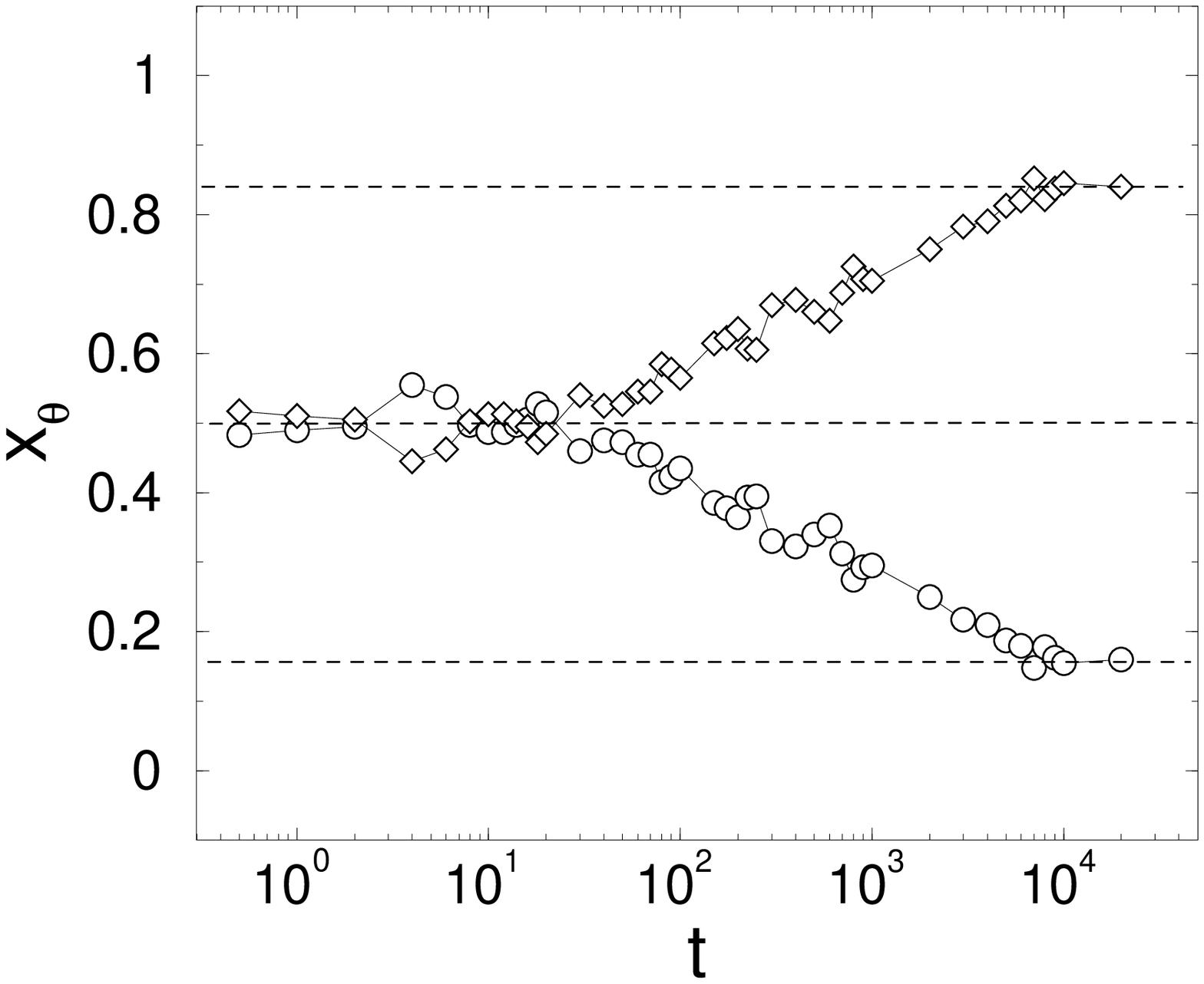,width=7.5cm}}
\caption[fig4]{
  Relative subpopulation sizes $x_{+}$ ($\Diamond$) and $x_{-}$
  ($\circ$) vs. time $t$ for the computer simulation shown in
  \pic{snap}. \label{w0-w1-06}}
\end{figure} 

As the different snapshots of \pic{snap} show, the minority and majority
organize themselve in space in such a way that they are separated. Thus,
besides the existence of a global majority, we find regions in the system
which are dominated by the minority. From this we can conclude a
\emph{spatial coordination of decisions}, i.e. agents which share the
same opinion are spatially concentrated in particular regions. With
respect to the example of the recycling campaign this means that those
agents who defect to cooperate (or cooperate in the opposite case),
are mostly found in a spatial domain of a like-minded neighborhood.  This
result might remind on the famous simulations of segregation in social
systems \citep{schelling-69,sakoda-71,hegselmann-flache-98} -
however, we would like to note that in our case the agents do \emph{not
  migrate} toward supportive places; they rather \emph{adapt} to the
opinion of their neighborhood based on the information received. 

The spatial distribution of the majority and the minority is also
reflected in the different components of the communication field, as
shown in \pic{hp-hm}. We find that the maxima of both components are of
about equal value, however, the information generated by the majority, is
roughly spread over the whole system, whereas the information
generated by the minority eventually concentrates only in specific
regions dominated by them. 
\begin{figure}[ht] 
\centerline{
\psfig{figure=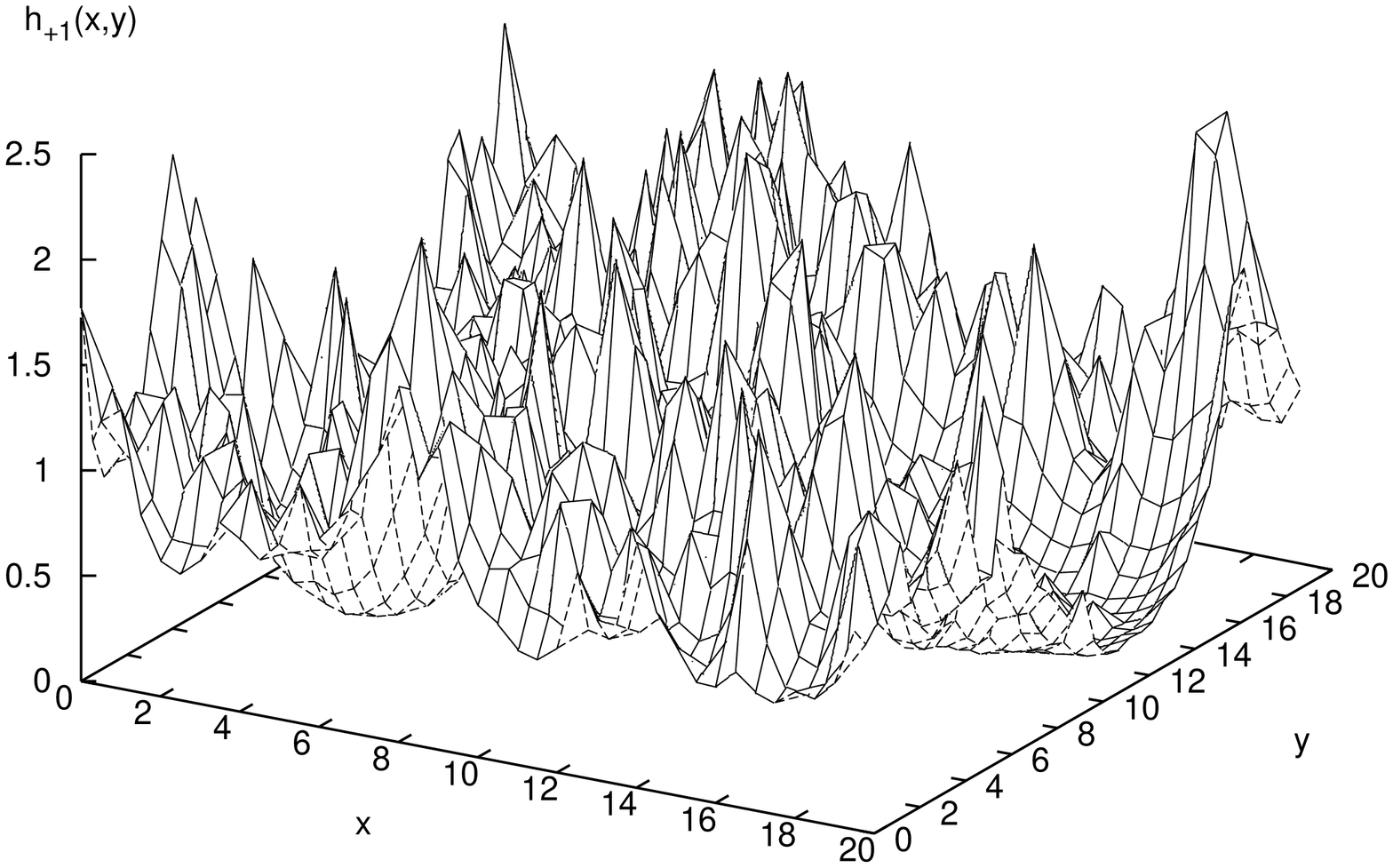,width=8.5cm,angle=-90}} 
\centerline{
\psfig{figure=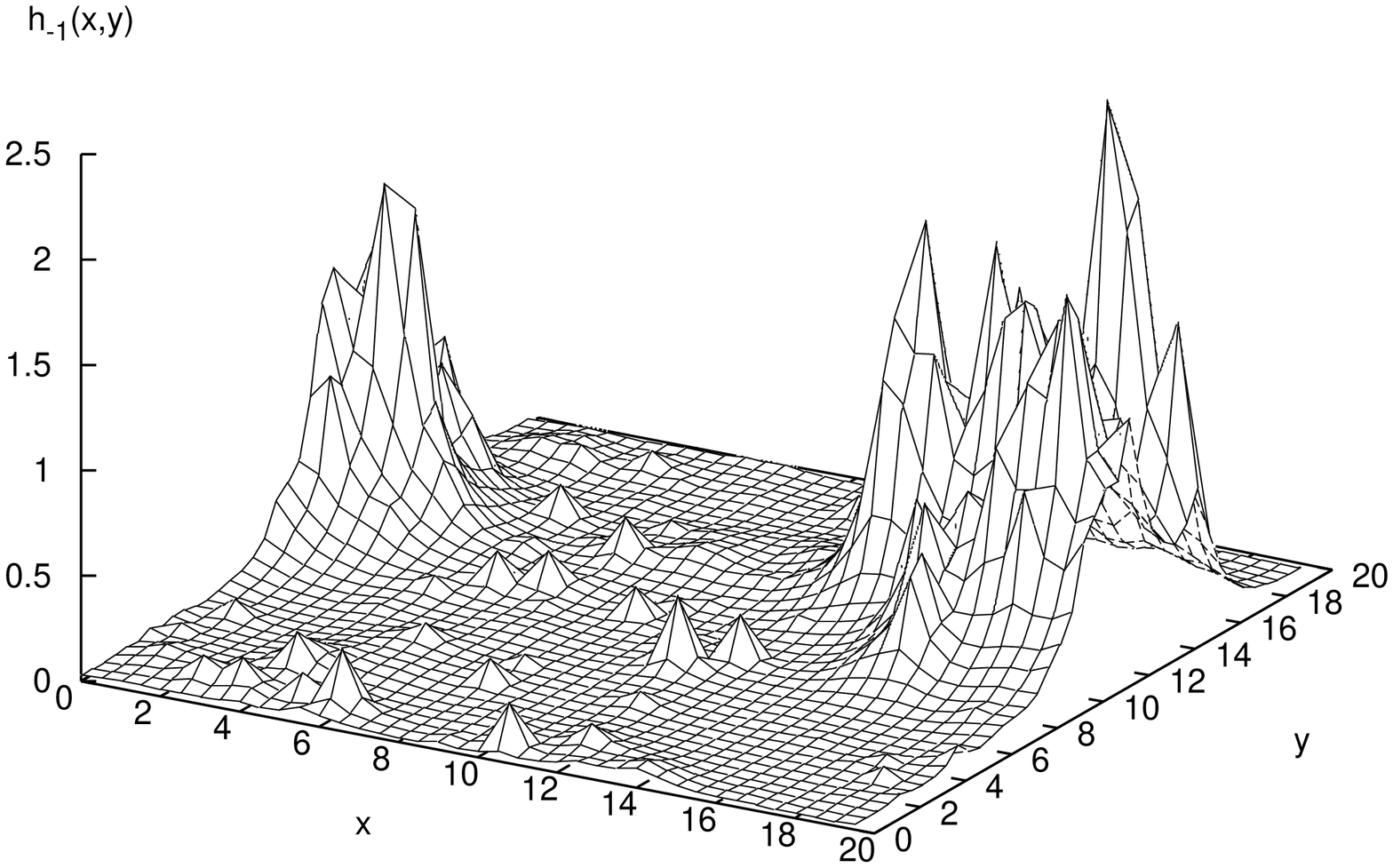,width=8.5cm,angle=-90} 
} 
\caption[fig2]{Spatial distribution of the two-component communication
  field: (top) $h_{+}(\bbox{r},t)$, (bottom)  $h_{-}(\bbox{r},t)$ at
  time $t=10^{4}$, which refers to the spatial agent distribution of
  \pic{snap}c. 
\label{hp-hm}} 
\end{figure} 

Running the simulations of the agent system different times with the same
set of parameters but just different initial random seeds, reveals an
interesting effect that can be observed in \pic{traj}, in comparison to
\pic{w0-w1-06}. We see that instead of a single fixed majority/minority
relation for the spatially extended system a \emph{large range} of such
relations exist, which even in the presence of fluctuations are stable
over a very long time. This means on the other hand that under certain
conditions the global outcome of the decision process becomes \emph{hard to
predict}. 

The different global majority/minority relations further 
correspond to \emph{different spatial coordination patterns}.
\pic{snap1600} gives a snapshot of the spatial distribution of
collaborators and defectors of a much larger system, but with the same
parameters as in \pic{snap} (in particular with the same average density
of agents).  We see that in this case the majority and the minority are
of about the same size. Further, the minority no longer exists rather
isolated (cf. \pic{snap}c), but in a spatially extended domain. We note
that besides some stochastic fluctuations the observed coordination
pattern remains stable also in the long run ($t=5\cdot10^{4}$).
\begin{figure}[htbp]
\centerline{\psfig{file=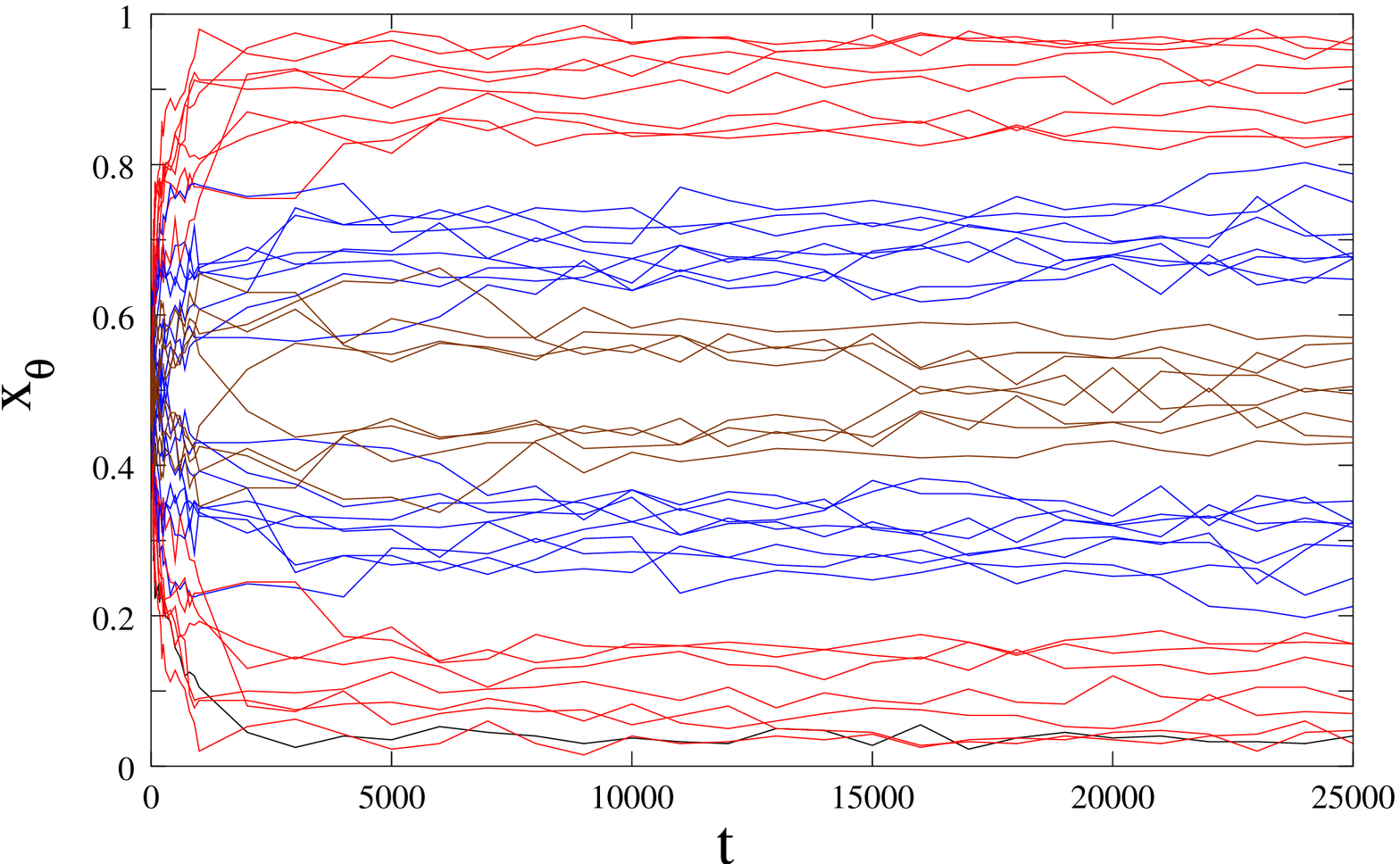,width=12cm}}
    \caption{Relative subpopulation sizes $x_{\theta}$ vs. time $t$ from
      20 computer simulations with the same set of parameters as in
      \pic{snap}, but different initial random seeds.}
    \label{traj}
\end{figure}
\begin{figure}[htbp]
\centerline{
\psfig{figure=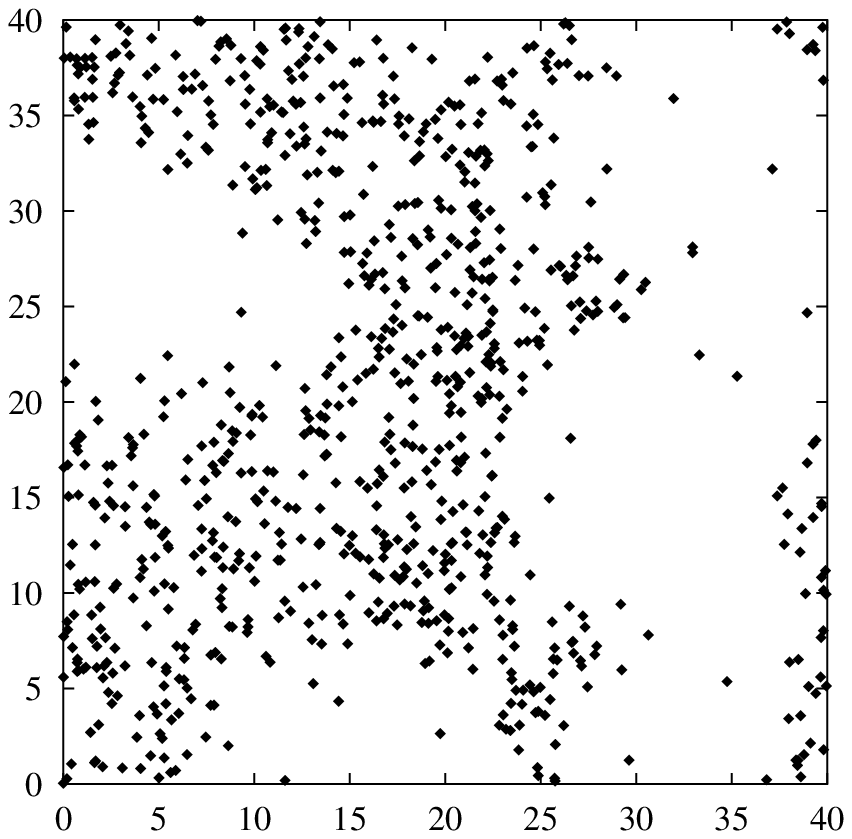,width=6.5cm}
\hfill
\psfig{figure=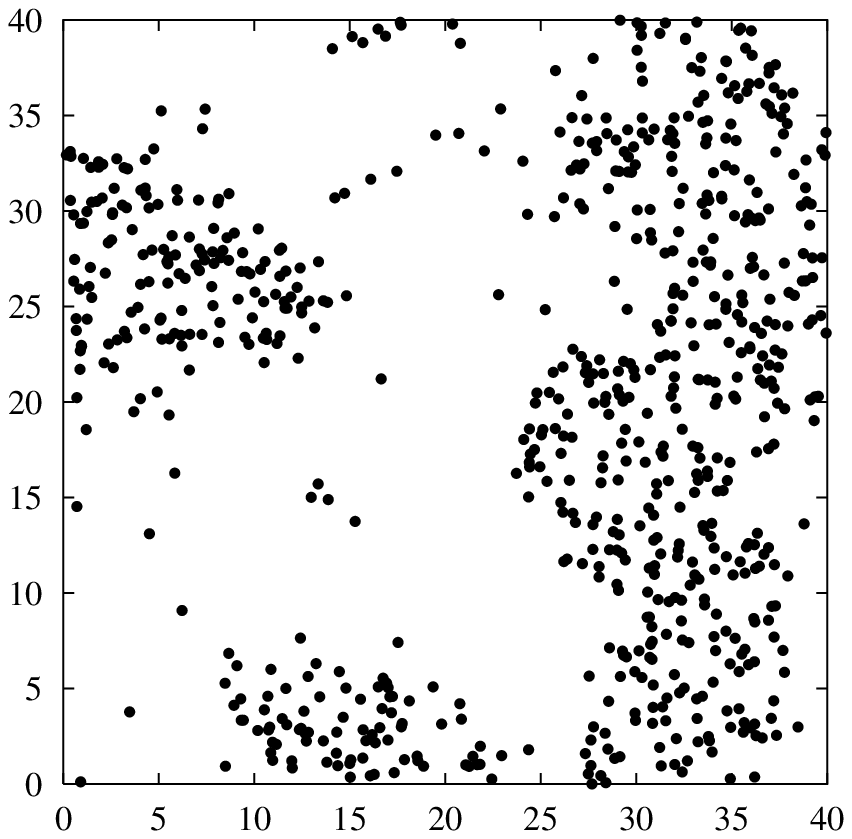,width=6.5cm}}
\caption[fig3]{
  Snapshot of the spatial distribution of cooperators ($\Diamond$ left)
  and defectors ($\circ$ right) at $t=5\cdot10^{4}$.  System size
  $A=1600$, total number of agents $N=1600$, for the other parameters see
  \pic{snap}. In this particular realization, the frequency of
  collaborators is $x_{+}=0.543$ and the frequency of defectors is
  $x_{-}=0.456$, respectively.
\label{snap1600}} 
\end{figure} 

Thus, we may conclude from these simulations that the spatially extended
system -- under certain conditions -- possesses multiple attractors for
the collective dynamics that makes the outcome of the decision process
hard to predict. This holds not only for the global minority/majority
ratio, but also for the possible spatial patterns that correspond to the
different attractors. This shall be discussed in more detail in the
following section. 

\subsection{Analytical Investigation of the Two-Box Case}
\label{4.2}

In order to get more insight into the attractor structure of the
spatially extended system, we want to investigate analytically the most
simple spatial case, consisting of only \emph{two boxes}. The frequency
to find agents with opinion $+1$ in box $i\in \{1,2\}$ shall be denoted
by $y_{i+}$, whereas the frequency in the total system is still given by
$x_{+}=N_{+}/N$, \eqn{fraction}. With a total system area $A$ and an
assumed homogeneous distribution of agents with an average density
$\bar{n}=N/A$, it yields:
\begin{eqnarray}
  \label{yplus}
  y_{i+}&=&\frac{N_{i+}}{N/2}\,=\,1-y_{i-}\;;\quad i=1,2 \\
x_{+}&=&\frac{1}{2}\left(y_{1+}+y_{2+}\right)
\end{eqnarray}
\begin{figure}[htbp]
\centerline{\psfig{figure=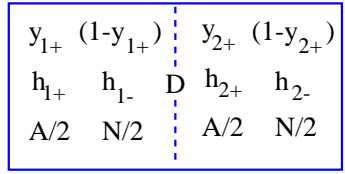,width=4.5cm}}
\caption[]{Sketch of the two-box system indicating the different
  variables. 
\label{2box}} 
\end{figure} 

\pic{2box} gives an overview of the relevant variables of the two-box
system.  The dynamics is described by six coupled equations, which read
explicitely:
\begin{eqnarray}
  \label{2box-set}
  \dot{y}_{i+}&=&(1-y_{i+})\,\eta
  \exp{\left[\frac{h_{i+}-h_{i-}}{T}\right]}
+ y_{i+}\,\eta
  \exp{\left[-\frac{h_{i+}-h_{i-}}{T}\right]}
 \;; \quad i=1,2 \nonumber \\
\dot{h}_{i+}&=&-k_{+}h_{i+}+s_{+}\bar{n}y_{i+}+D_{+}\frac{
8\,(h_{j+}-h_{i+})}{A} \;;\quad i,j\in \{1,2\}\;;\;\, i\neq j
  \nonumber \\
\dot{h}_{i-}&=&-k_{-}h_{i-}+s_{-}\bar{n}(1-y_{i+}) +D_{-}\frac{
8\,(h_{j-}-h_{i-})}{A} 
\end{eqnarray}
The stationary solutions of the set of equations follow from
$\dot{y}_{i+}=0$; $\dot{h}_{i+}=\dot{h}_{i-}=0$.  Combining the six
equations of \eqn{2box-set} in the stationary case, we arrive after some
transformations at the following two conditions:
\begin{multline}
  \label{2b-2a}
   \ln{\left(\frac{y_{1+}}{1-y_{1+}}\right)}+
  \ln{\left(\frac{y_{2+}}{1-y_{2+}}\right)}
 =   2 \Big(y_{1+}+y_{2+}\Big) \left[\frac{s_{+}\bar{n}}{k_{+}T} +
\frac{s_{-}\bar{n}}{k_{-}T}
\right] - 4\frac{s_{-}\bar{n}}{k_{-}T} \\
\shoveleft{\ln{\left(\frac{y_{1+}}{1-y_{1+}}\right)}
-  \ln{\left(\frac{y_{2+}}{1-y_{2+}}\right)}
= 
2 \Big(y_{1+}-y_{2+}\Big) \times} \\
\times \left[\frac{s_{+}\bar{n}}{k_{+}T} 
\left(1+\frac{16 D_{+}}{A k_{+}}\right)^{-1}
+ \frac{s_{-}\bar{n}}{k_{-}T}
\left(1+\frac{16 D_{-}}{A k_{-}}\right)^{-1}\right]
\end{multline}
The coupled equations of \eqn{2b-2a} define the possible stationary
frequencies $y_{1+}$, $y_{2+}$ of agents in the two coupled boxes and
therefore serve as a starting point for a subsequent bifurcation
analysis. It will be convenient to assume first the parameters of the
different field components as equal again, cf. \eqn{equal}. In this case,
we can recover the bifurcation parameter $\kappa$, \eqn{a-stat} of the
mean field limit as follows:
\begin{eqnarray}
  \label{k1-k2}
\kappa & = & 
\frac{s_{+}\bar{n}}{k_{+}T} + \frac{s_{-}\bar{n}}{k_{-}T} 
= \frac{2\, s\,\bar{n}}{k\,T}
\end{eqnarray}
By means of the previously defined function $\mathcal{L}(x,\lambda)$,
\eqn{l-xl-def} we are now able to express the conditions for stationary
solutions, \eqn{2b-2a} in the following compact form:  
\begin{eqnarray}
  \label{2b-comp}
\mathcal{F}_{1+}&:=&\kappa(1-\bar{D})\left(y_{2+}-y_{1+}\right) 
+ \mathcal{L}(y_{2+},2\kappa) = 0\nonumber \\
\mathcal{F}_{2+}&:=&\kappa(1-\bar{D})\left(y_{1+}-y_{2+}\right)  
+\mathcal{L}(y_{1+},2\kappa)=0
\end{eqnarray}
where 
\begin{equation}
\label{bar-d}  
\bar{D}=\left(1+\frac{16 D}{A k}\right)^{-1}  
\end{equation}
If we compare \eqn{2b-comp} with \eqn{xl-stat} for the mean-field case,
we find that each solution $x_{+}$ of the mean-field case is also a
solution of \eqn{2b-comp} by setting $y_{1+}=y_{2+}=x_{+}$.  Moreover, if
$D=0$, i.e. no exchange between the two boxes, then $\bar{D}=1$, and
\eqn{2b-comp} reduces to \eqn{xl-stat}.

\begin{figure}[htbp]
\centerline{\psfig{file=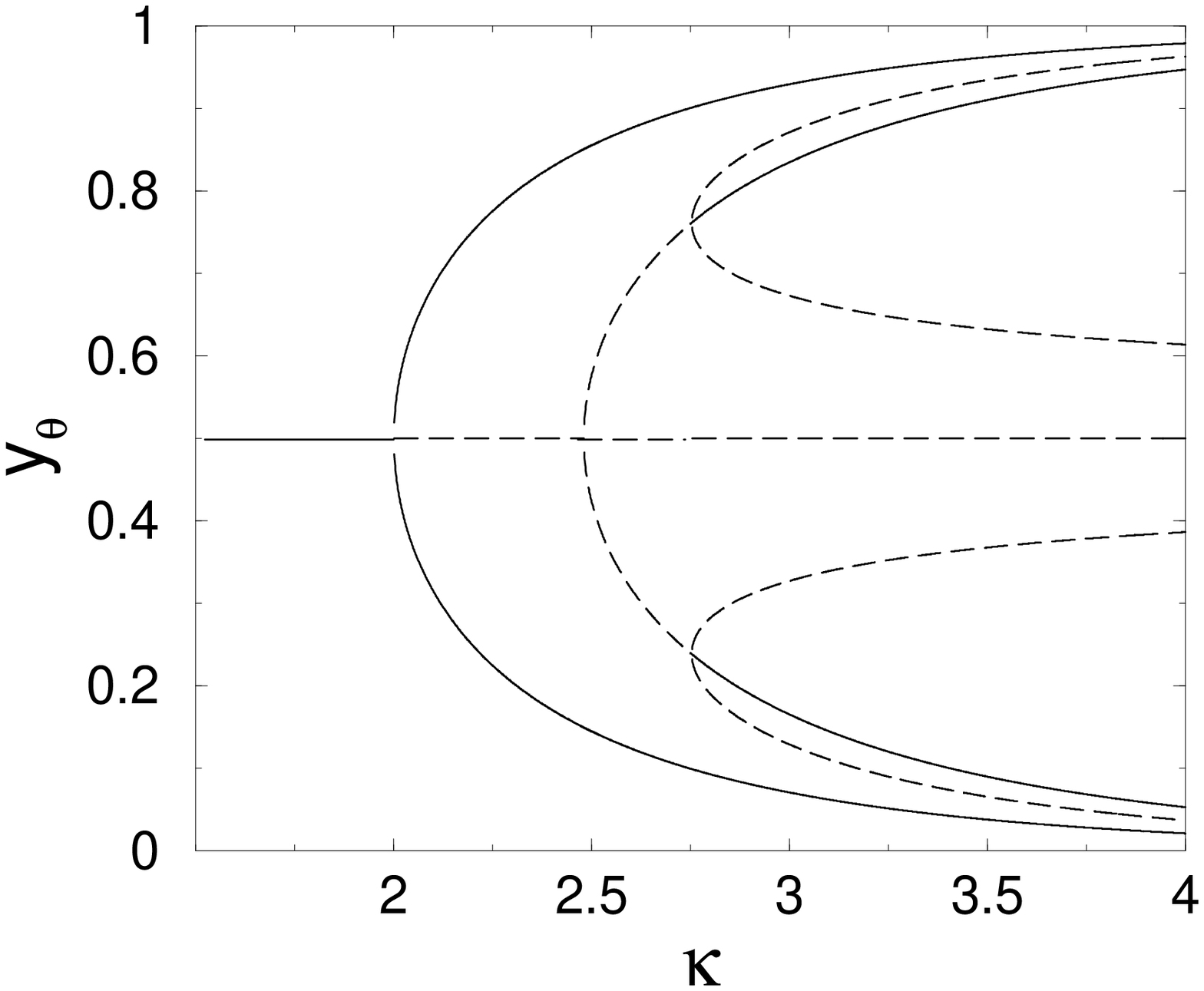,width=7.5cm}}
\centerline{\psfig{file=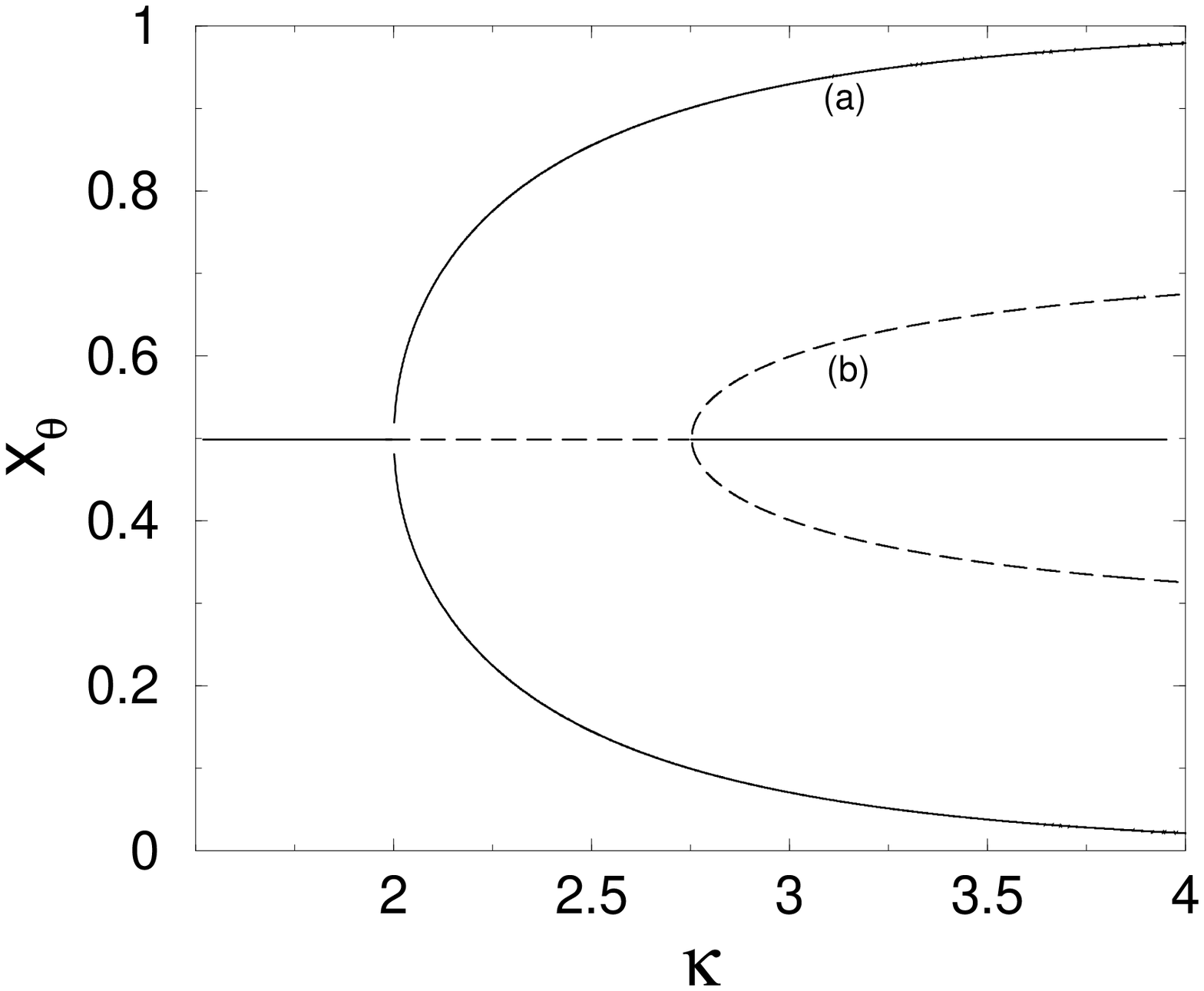,width=7.5cm}}
    \caption{Stationary solutions for $y_{\theta}$ (top: single box) and
      $x_{\theta}$ (bottom: total system) resulting from 
      \eqn{2b-comp} for different values of $\kappa$. Parameters:
      $A=400$, $D/k=6.0$. The solid lines indicate stable solutions, the
      dashed lines instable ones.
    \label{2bifurc}}
\end{figure}
The coupled equations (\ref{2b-comp}) have been solved numerically,
further we have conducted a stability analysis. The
corresponding bifurcation diagram is shown in \pic{2bifurc} for a single
box (top) and the total system (bottom).  If we compare the latter one
with the bifurcation diagram of the mean field limit, \pic{crit}, we find
that the bifurcation at $\kappa^{c}=2$ is obtained again, cf. curve (a).
That means, for $\kappa<\kappa^{c}$, the equal distribution of both
opinions is the stable state also for the spatially heterogeneous system,
while for $\kappa>\kappa^{c}$ a majority and a minority emerges, which
organizes itself in space in the way shown in \pic{snap}.

With increasing $\kappa$, we find a second bifurcation,
curve (b) at the critical value $\kappa^{c}_{2}$. For consistency, the
first critical value shall be denoted as
$\kappa_{1}^{c}\equiv\kappa^{c}=2$.  As long as
$\kappa_{1}^{c}<\kappa<\kappa_{2}^{c}$, the minority/majority ratio is
clearly defined by just one possible value, cf. \pic{2bifurc}(bottom).
For $\kappa_{2}^{c}<\kappa$, however there are different stable
majority/minority relations possible in the system.
This can be also clearly observed in \pic{2attrac} that shows the stable
and instable stationary states of the 2-box system for three different
values of $\kappa$.

\begin{figure}[htbp]
\centerline{\psfig{file=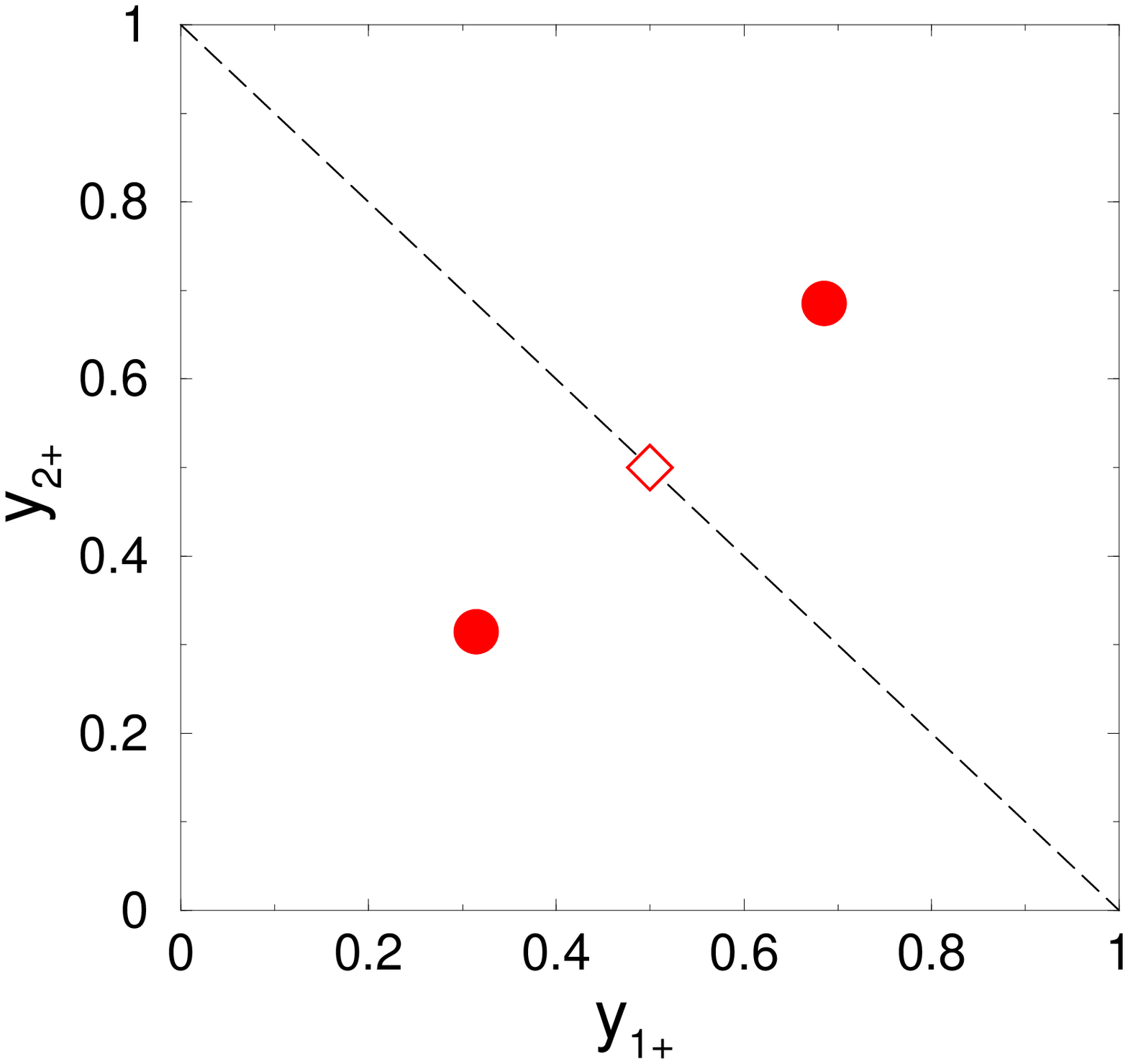,width=4.2cm}\hfill
\psfig{file=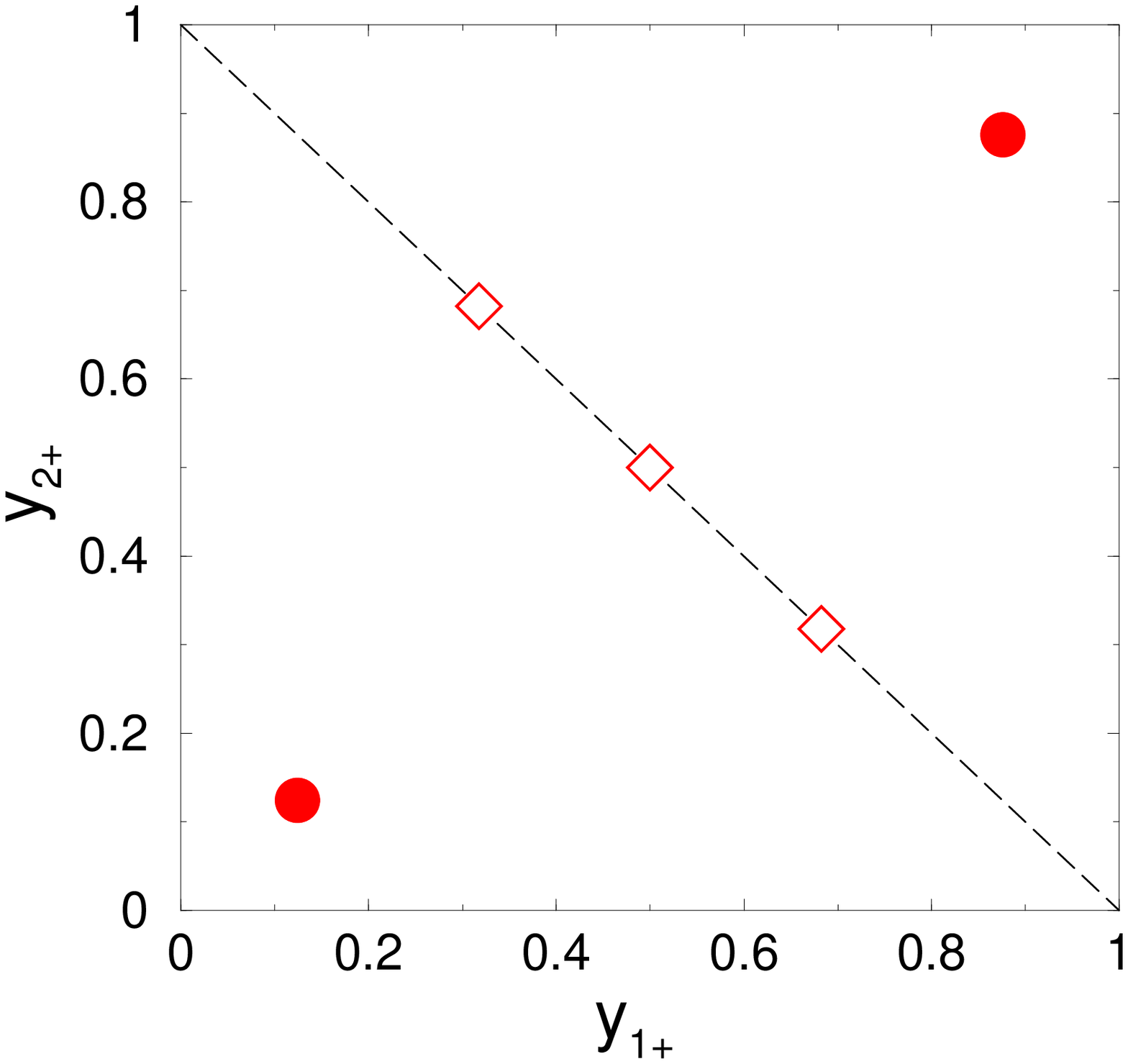,width=4.2cm}\hfill
\psfig{file=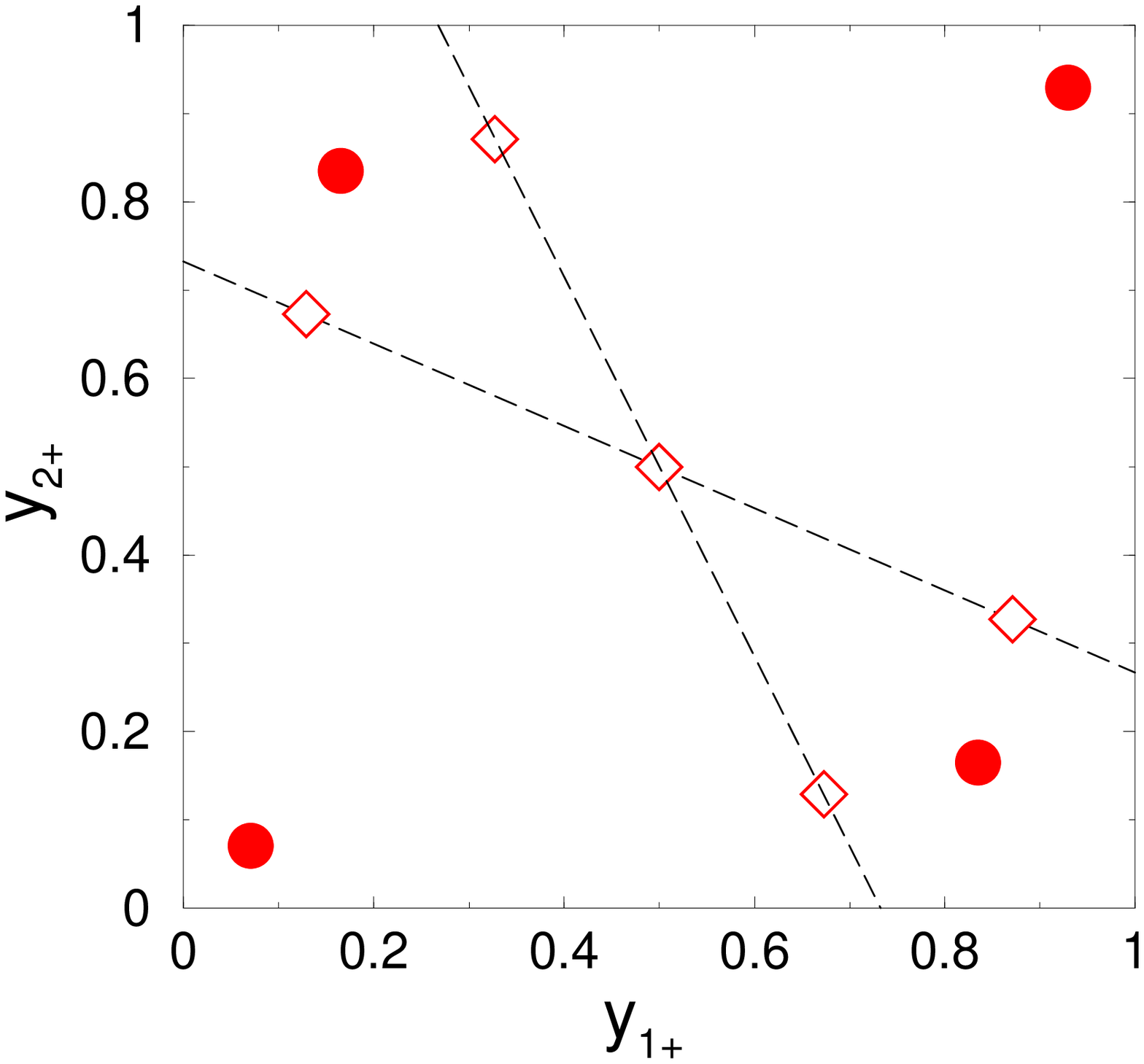,width=4.2cm}
}
    \caption{Stationary solutions for $y_{1+}$ vs. $y_{2+}$ resulting from 
      \eqn{2b-comp}. (left) $\kappa^{\prime}_{1}<\kappa=2.1 <\kappa^{\prime}_{2}$, (middle) $\kappa_{2}^{\prime}<\kappa=2.6<\kappa^{\prime}_{3}$, (right)
      $\kappa^{\prime}_{3}<\kappa=3.0$. 
 Other parameters see \pic{2bifurc}. The filled circles
      mark stable states, the open squares instable ones. The
      dashed lines are used to indicate possible separatrices. 
    \label{2attrac}}
\end{figure}

\begin{figure}[htbp]
\centerline{\psfig{file=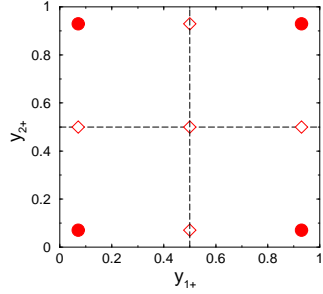,width=4.2cm}}
    \caption{Stationary solutions for $y_{1+}$ vs. $y_{2+}$ resulting from 
      \eqn{2b-comp} for 
      $\kappa$=3.0 and $D/k$=0, i.e. $\bar{D}$=1. This figure should be
      compared to \pic{2attrac}(right), where $D/k$=6.0.
    \label{0attrac}}
\end{figure}

The appearence of the second bifurcation curve (b) in
\pic{2bifurc}(bottom) can be understood when looking at the bifurcation
diagram of a single box at the top part of \pic{2bifurc}. To avoid
confusion, the critical bifurcation values of the \emph{single box} are
indicated by $^{\prime}$ instead of $^{c}$.  In the single box, the first
bifurcation is found at $\kappa_{1}^{\prime}=\kappa_{1}^{c}=2$ as in the
mean-field case, and a second bifurcation at
$\kappa_{2}^{\prime}>\kappa_{1}^{c}$. The resulting new stationary
states, however still lead to an average frequency $x_{\theta}=0.5$ in
the total system, because of $y_{1+}=1-y_{2+}$ in this particular case.
Only after a third bifurcation in the single box that occurs at the
critical value $\kappa_{3}^{\prime}=\kappa_{2}^{c}$, we find stationary
frequencies in each box that do not automatically sum up to 1, and
therefore allow for different frequencies for the total system.  That
means, in order to know about the possible appearence of different
majority/minority relations in the 2-box system, we have to concentrate
on the appearence of the third bifurcation for the single box.

Eventually, we also investigate how the different critical bifurcation
values $\kappa_{i}^{\prime}$ shift with the scaled diffusion constant
$D/k$. The numerical results are shown in \pic{bif}. 
While $\kappa_{1}^{\prime}$
remains constant, $\kappa_{2}^{\prime}$ increases linearly with $D/k$, and
$\kappa_{3}^{\prime}$ increases in 
nearly linear manner.
\begin{figure}[htbp]
  \centerline{\psfig{file=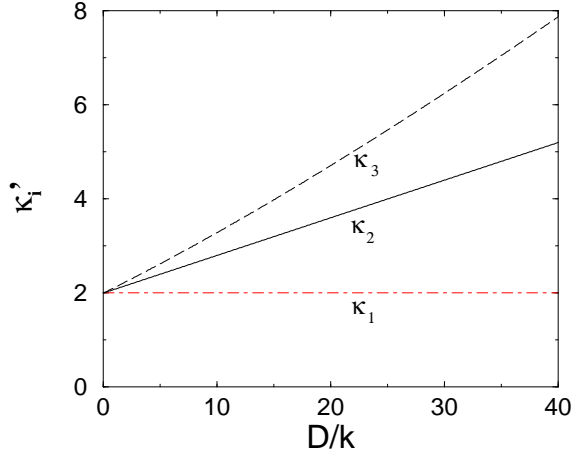,width=7.5cm}}
  \caption{Change of the critical bifurcation values $\kappa_{i}^{\prime}$
    (i=1,2,3) for a single box dependent on $D/k$ (cf. also \pic{2bifurc}
    top).}
  \label{bif}
\end{figure}

At least for the linear dependence, the shift of $\kappa^{\prime}$ with
$D/k$ can be derived analytically. As \pic{2bifurc}(top) shows, the first
two bifurcations occur for $y_{i+}=1/2$. Looking e.g. at box $i=1$, we
find for $\mathcal{F}_{1+}$, \eqn{2b-comp}:
$\mathcal{F}_{1+}\left(y_{1+}=1/2\right)=0$.  The dependence
$\kappa^{\prime}(\bar{D})$ results then from the condition:
\begin{equation}
  \label{f-df}
\left.\frac{\partial \mathcal{F}_{1+}}{\partial
    y_{1+}}\right|_{y_{1+}=1/2}=0
\end{equation}
This eventually leads to a quadratic equation for
$\kappa^{\prime}(\bar{D})$:
\begin{equation}
  \label{quadrat}
  \bar{D}\,{\kappa^{\prime}}^{2}-2(1+\bar{D})\,\kappa^{\prime}+4=0
\end{equation}
with the two solutions
\begin{equation}
  \label{q-sol}
  \kappa_{1}^{\prime}=2\;;\quad
  \kappa_{2}^{\prime}=\frac{2}{\bar{D}}=2\left(1+\frac{16 D}{A\,k}\right)
\end{equation}
which agree with the results of \pic{bif}. 

We note that all three bifurcations keep existing also in the limit
$D/k\to 0$, until the bifurcation values $\kappa_{i}^{\prime}$ coincide
at $D/k = 0$. The diffusion between the two boxes of course affects the
values of the $\kappa_{i}^{\prime}$, as shown in \pic{bif}, but the
values of the \emph{stable} states are not changed much. This can be seen
by comparing \pic{2attrac}(right) with \pic{0attrac} valid for $D/k = 0$,
that clearly indicates how the \emph{size of the attractor regions}
around the stable states is influenced by diffusion. In the limit of $D/k
\to \infty$ the situation of the mean-field limit is recovered again. 

In conclusion, the investigation of the two-box case has shown that above
a certain critical value $\kappa_{2}^{c}=\kappa_{3}^{\prime}$ the
possibility of \emph{different} stable majority/minority ratios exist.
For a given supercritical population density, cf. \eqn{n-crit}, the
velocity of information exchange, expressed in terms of the diffusion
constant $D$ and the life time of information $k$ determine whether
these multiple attractors can exist.

These results can be also adapted to the spatially extended ``multi-box''
case as an approximation of the continuous space. If for the single box
multiple attractors exist, we can observe in the multi-box case a larger
variety of spatial coordination patters, with different patches of
likeminded agents coexisting (cf.  \pic{snap1600}). Regarding the
possible values of the minority/majority ratio for the total system, we
note that in the spatially extended system almost every value can be
realized. These \emph{multiple} stable states for the total system can be
envisioned as a combination of the different possible stable states of
the single boxes.  An increase of the diffusion constant of the
information field, on the other hand, restricts the stability of multiple
attractors of the single boxes and this way limitates the variety of
possible spatial coordination patterns.

\section{Information Dissemination on Different Time Scales}
\label{5}

From the investigations in the previous section, we found the emergence
of different majority/minority relations in a spatially extended agent
system. So far, however, fluctuations during the initial time lag
observed may decide whether the cooperators or the defectors will appear
as the majority. If we start from an unbiased initial distribution, i.e.
an equal distribution between both opinions, then there is no easy way to
break the symmetry towards e.g. cooperation, except, an external bias is
taken into account.

In this section, we will investigate a possibility to break the symmetry
by means of different information dissemination, i.e. we may exploit the
different properties of the information exchange in the system, as
expressed in terms of the parameters $s_{\theta}$, $k_{\theta}$,
$D_{\theta}$ of the communication field. For instance, we may assume that
the information generated by of one of the subpopulations is distributed
\emph{faster} in the system than the information generated by the other
one. Alternatively, we may also consider different life times of the
different components of the communication field.  However, in order to
model a faster exchange of information, it is not sufficient to simply
increase the value of $D_{\theta}$, we need to consider its effect on the
local values of the communication field in more detail.
 
\Eqn{hrt} is known as a reaction-diffusion equation. In order to
elucidate the influence of the different parameters, we will assume for
the moment a one-dimensional system with the spatial coordinate $x$.
Further, there shall be only one group of $N$ agents, $\theta=0$, which
are located at the same place, $x=0$ and have the same personal strength,
$s_{0}$. Then, \eqn{hrt} can be simplified to:
\begin{equation}
  \label{hxt}
\frac{\partial}{\partial t} h_{0}(x,t) = 
\frac{s_{0}\,N}{A}\, \delta(x-0) - \;k_{0}\,h_{0}(x,t) \;+ \;D_{0}\,
\Delta h_{0}(x,t) 
\end{equation}
Initially, $h(x,0)=0$ yields.  After a certain time which depends in
particular on $D_{0}$, the communication field $h_{0}(x,t)$ generated by
the agents will reach a stationary distribution. With $\dot{h}(x,t)=0$
and $n=N/A$ we find from \eqn{hxt}:
\begin{equation}
  \label{hr-stat}
  h_{0}^{stat}(x)= \frac{s_{0} n}{2\, \sqrt{k_{0}\,D_{0}}}  \; \exp{\left\{- \sqrt{\frac{k_{0}}{D_{0}}}\, \abs{x}\right\}}
\end{equation}
\begin{figure}[ht] 
\centerline{\psfig{figure=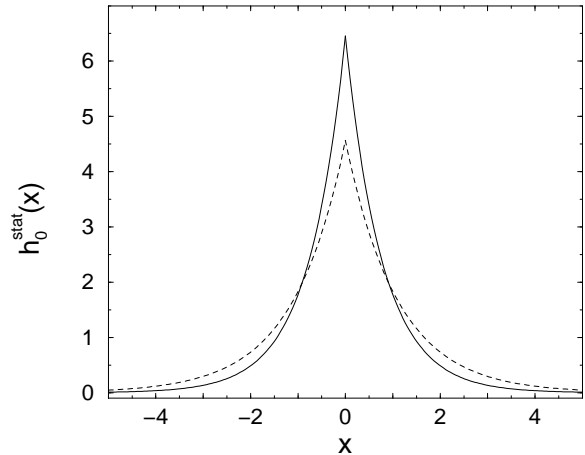,width=7.5cm,angle=-90}} 
\caption[fig2]{
  Stationary distribution $h_{0}^{stat}(x)$, \eqn{hr-stat}, of the
  communication field, assumed that there is only one group of $N$ agents
  located at the same place, $x=0$. $(-\!\!\!-\!\!\!-\!\!\!-)$
  $D_{0}=0.06$, $(-\,-)$ $D_{0}=0.12$, other parameters: $N=10$, $A=1$,
  $s_{0}=0.1$, $k_{0}=0.1$.
\label{hr-stat-pic}} 
\end{figure} 

\Eqn{hr-stat} is plotted in \pic{hr-stat-pic} for two different
parameters of the diffusion constant $D_{0}$. Obviously, in addition to
the time scale of information dissemination, an increase of $D_{0}$ mainly
affects the stationary value of the communication field $h_{0}^{stat}(x)$
at the position of the agents, $x=0$. I.e. a faster communication in the
system via a faster diffusion of the generated information, also lowers
the information available at the agent's position. This might be
considered as a drawback in modeling information exchange by means of
reaction-diffusion equations. Obviously, the field
$h_{\theta}(\bbox{r},t)$ obeys certain boundary conditions and
conservation laws, which do not hold for ``information \emph{per se}''. 
In particular, the local value of available information is not lowered if
this information spreads out faster, but the local value of the
``communication field'' obeying \eqn{hrt} does.

In order to compensate the unwanted effect of a local decrease of
$h_{\theta}(\bbox{r},t)$, the production rate $s_{\theta}$ can be
increased accordingly, which however effects again the stationary values. 
To be consistent, we have to choose that both the ratios
\begin{equation}
  \label{ratio}
 \frac{k_{\theta}}{s_{\theta}}=\beta;; \quad 
 \frac{D_{\theta}}{s_{\theta}}=\gamma
\end{equation}
need to be constant for both components $\theta=\{+1,-1\}$. In this case,
\eqn{hrt} for the dynamics of the multi-component communication field can
be rewritten as:
\begin{equation} 
\label{hrt-red} 
\frac{\partial}{\partial \tau} h_{\theta}(\bbox{r},\tau) = 
\sum_{i=1}^{N}\delta_{\theta,\theta_{i}}\;
\delta(\bbox{r}-\bbox{r}_{i})\;
- \;\beta\; h_{\theta}(\bbox{r},\tau) \;+ \;\gamma\;
\Delta h_{\theta}(\bbox{r},\tau). 
\end{equation} 
where the time scale $\tau$ is now defined as
$\tau=t\,(D_{\theta}/\gamma)$. If both parameters $\beta$
and $\gamma$ are kept constant, \eqn{hrt-red} means that the dynamics of
the respective component of the communication field occurs on a different
time scale $\tau$, dependent on the value of $D_{\theta}$. An increase in
the diffusion constant $D_{\theta}$ then models indeed the information
exchange on a faster time scale as expected, without affecting the
stationary distribution resulting from \eqn{hrt-red}.

The effect of the different diffusion constants can be understood by
means of computer simulations, where the parameters $\beta$ and $\gamma$,
\eqn{ratio} are kept constant and only the ratio
\begin{equation}
  \label{d}
  d=\frac{D_{+}}{D_{-}}
\end{equation}
is varied. 
\pic{w0-w1-adj} shows the evolution of the subpopulations in time for
$d=1/3$ and $d=1$ for comparision. 
\begin{figure}[ht] 
\centerline{\psfig{figure=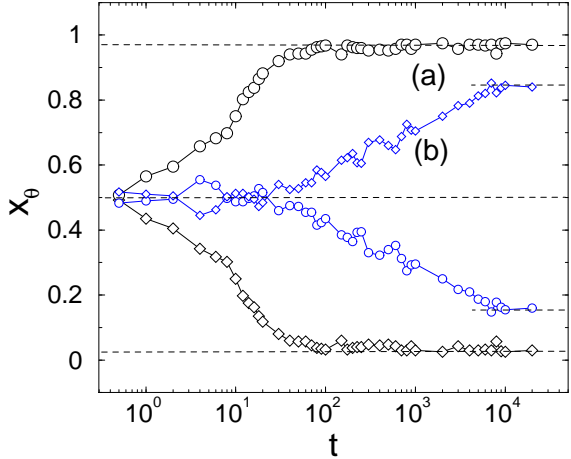,width=7.5cm}} 
\caption[fig2]{
  Relative subpopulation sizes $x_{+}$ ($\Diamond$) and $x_{-}$ ($\circ$)
  vs. time $t$. Parameters: $A=400$,
  $N=400$, $\beta=1$, $\gamma=0.6$, (a) $D_{+}=0.02$, $D_{-}=0.06$, (b)
  $D_{+}=D_{-}=0.06$ (cf. \pic{w0-w1-06}).
\label{w0-w1-adj}}
\end{figure} 
Two features can be noticed from these two particular runs: (i) For
$d=1/3$, the initial time lag when the decision which subpopulation
becomes the majority is yet pending, has vanished. I.e. compared to
$d=1$, there is a considerably reduced period of time for early
fluctuations to break the symmetry toward one of the subpopulations. The
time lag has been related before to the establishment of the
communication field needed to provide the coupling between the agents.
But here we see that the symmetry is broken very fast, even without a
fully established communication field.  (ii) For $d=1/3$, the
subpopulation with the faster (more ``efficient'') communication has
become the majority, while for $d=1$ both subpopulations have an equal
chance to become the majority in the system.  These conclusions however
need some substantiation, therefore we have conducted more extensive
computer simulations presented in the following.

\pic{d1d2-n} shows the total fraction of agents of subpopulation $\{-1\}$
over time averaged over 20 runs, for different values of $d$. This mean
value gives an estimate of the chance that subpopulation $\{-1\}$ becomes
the minority or majority in the system. We find that for $d=1$ this
chance is about 50 percent, i.e. only random events decide about its
status, as long as the information dissemination of both subpopulations
occurs on the same time scale. With increasing $d$ (i.e. with an
increasing information diffusion of the \emph{other} subpopulation), the
trend towards a minority status clearly increases for subpopulation
$\{-1\}$, as shown in the average fraction $\mean{x_{-}}$. Already for
$d=1.5$, i.e. for example $D_{+}=0.06$, $D_{-}=0.04$, the average
subpopulation fraction reaches a constant minimum value in the stationary
limit, which means that deviations in size have considerably decreased.
\begin{figure}[ht] 
\centerline{\psfig{figure=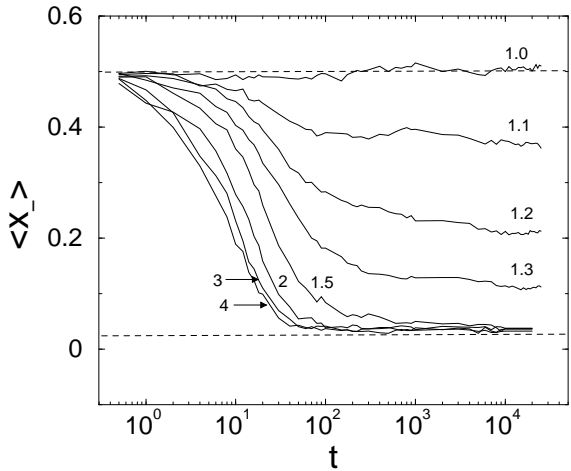,width=7.5cm}} 
\caption[]{
  Relative subpopulation size $\mean{x_{-}}$ averaged over 20 runs.  The
  different numbers give the value of $d=D_{+}/D_{-}$. Other parameters
  see \pic{w0-w1-adj}.
\label{d1d2-n}}
\end{figure} 

A closer inspection is given in the series of \pic{error}, which shows
 the mean values of $\mean{x_{-}}$ together with the minimum and maximum
 values of the 20 runs to indicate the scattering. 
\begin{figure}[ht] 
\centerline{(a)\hfill(b)}\bigskip
\centerline{\psfig{figure=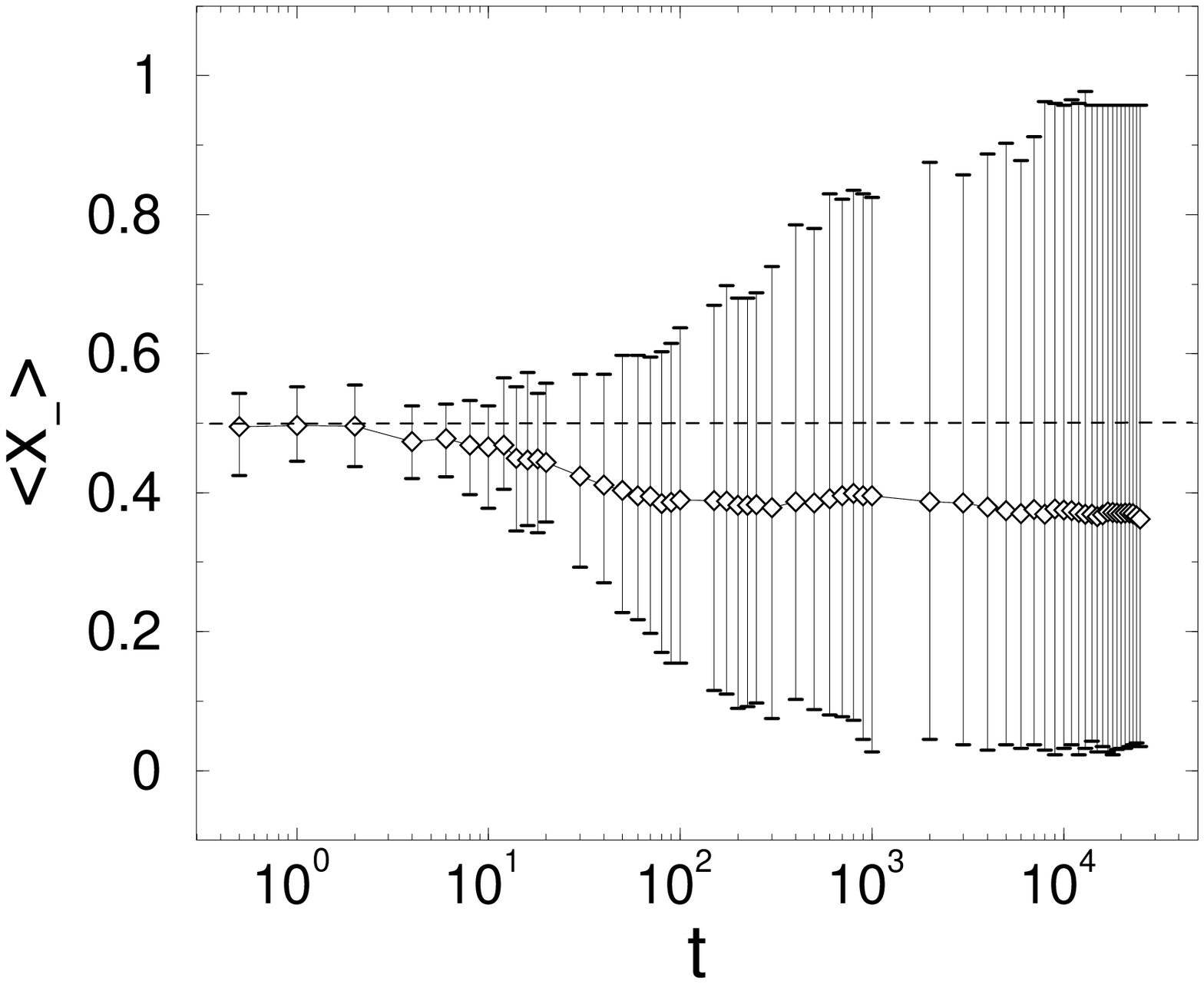,width=6.5cm}\hfill
\psfig{figure=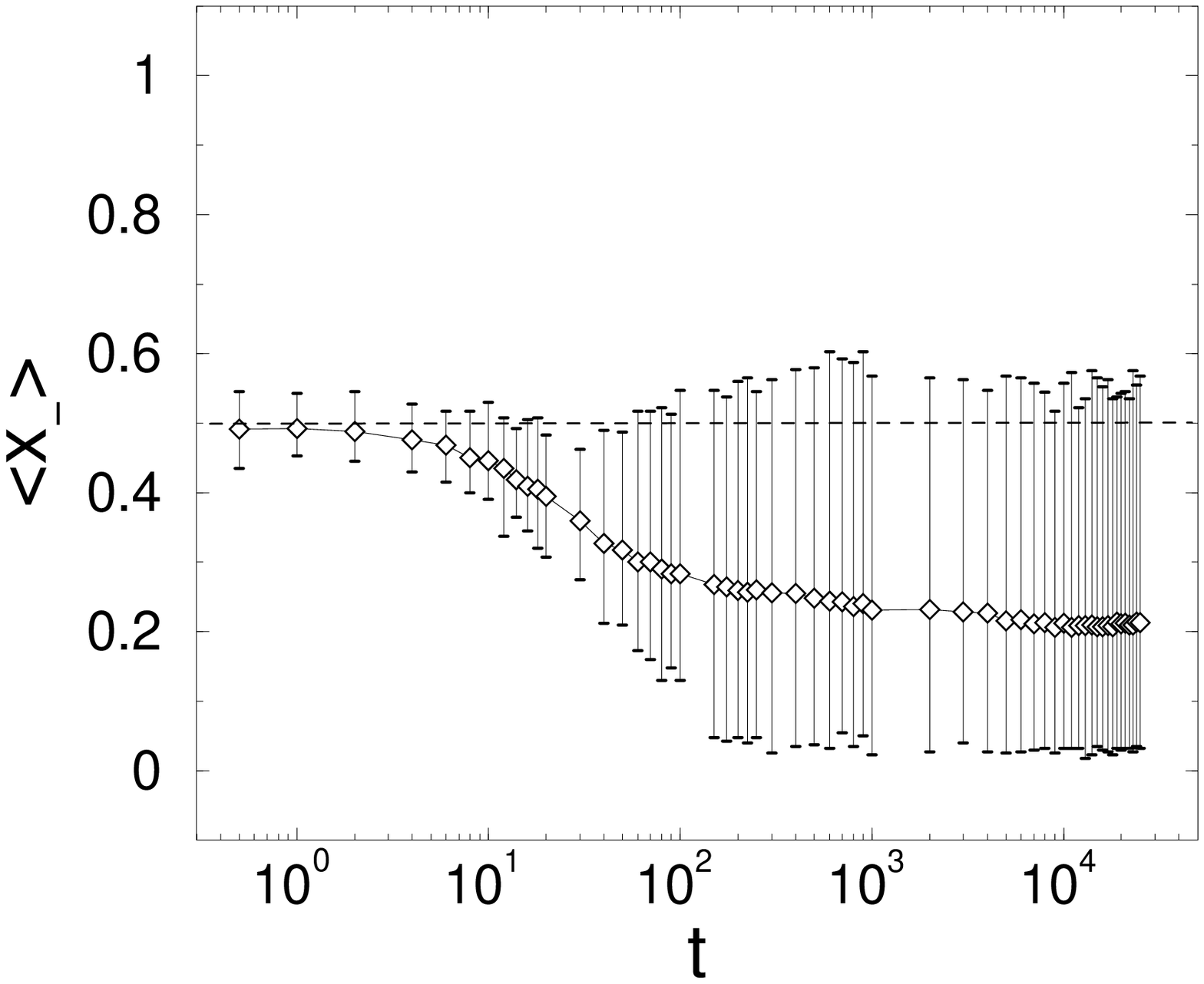,width=6.5cm}}

\centerline{(c)\hfill(d)}\bigskip
\centerline{\psfig{figure=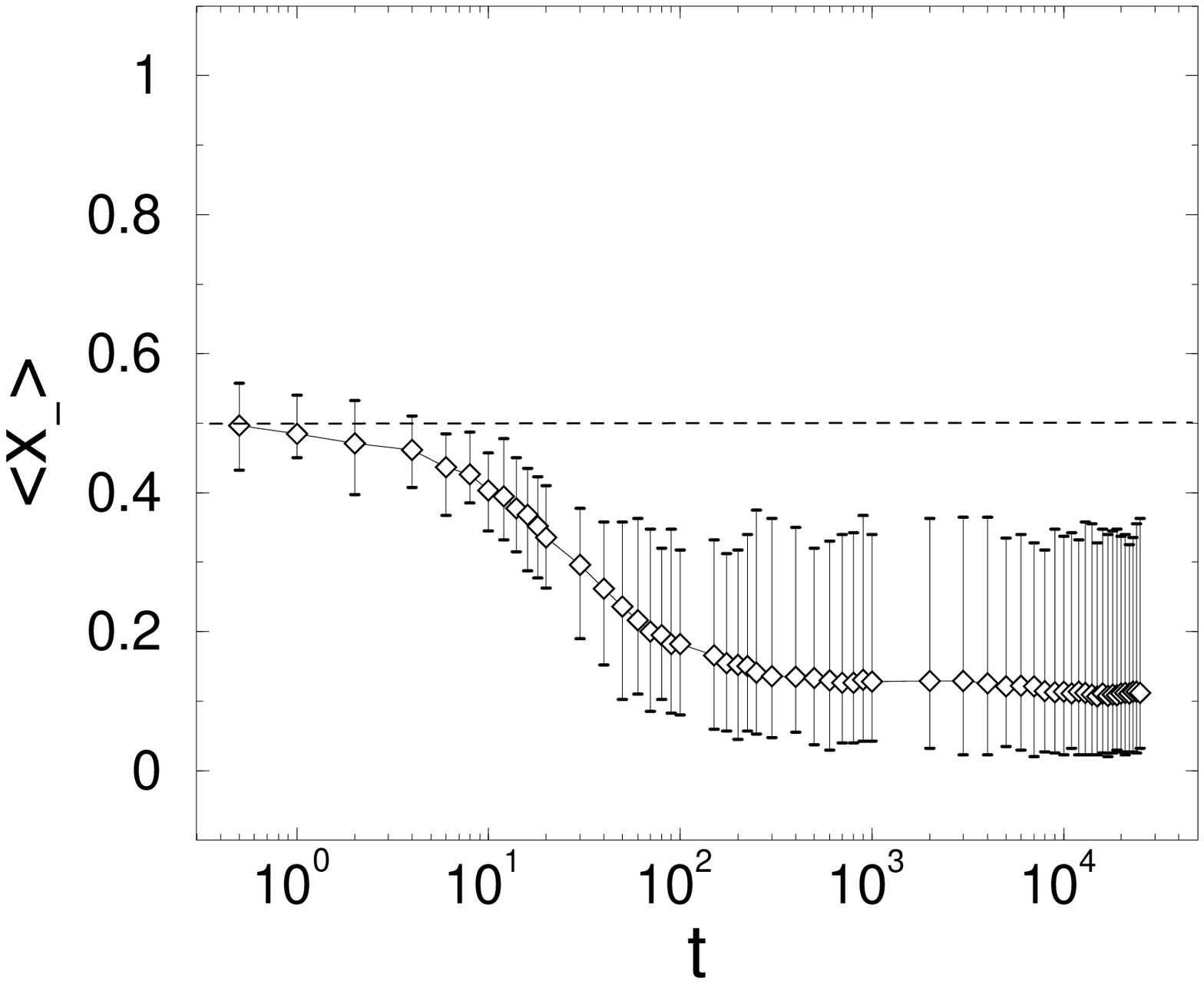,width=6.5cm}\hfill
\psfig{figure=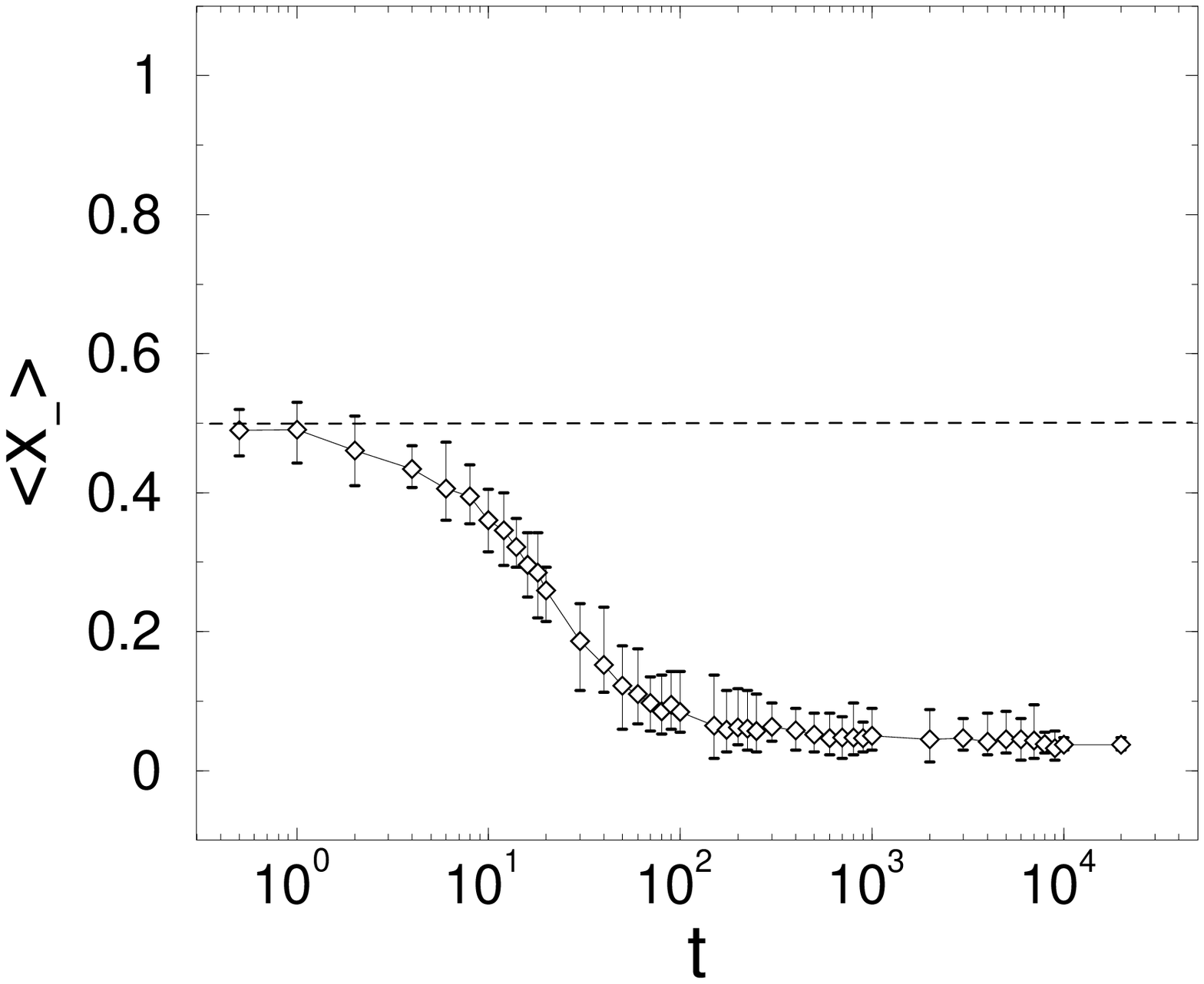,width=6.5cm}} 
\caption[fig2]{
  Relative subpopulation size $\mean{x_{-}}$ averaged over 20 runs (cf.
  \pic{d1d2-n}) together with minimum/maximum values. (a) $d=1.1$, (b)
  $d=1.2$, (c) $d=1.3$, (d) $d=1.5$. Other parameters see
  \pic{w0-w1-adj}.
\label{error}}
\end{figure} 

We find that for $d\in \{1.1;1.2\}$, inspite of the clear tendency
towards the minority status, there are still possibilities that the
subpopulation $\{-1\}$ ends up as the majority in the system -- even with
a slower communication. With increasing $d$, theses possibitities vanish,
as \pic{error} (c,d) indicate. However, for a range of $d\in\{1.2;1.4\}$
we find that the \emph{size} of the minority population still shows a
large range of possible values. Only for $d>1.5$, these deviations become
small enough to allow only one stable size of the minority subpopulation.

Considering the results for possible attractors in the spatially extended
system obtained in the previous section, we can discuss
the outcome of the above simulations also from a different perspective.
They allow to distinguish between two different stationary regimes: (i) a
\emph{multi-attractor} regime, where different values for a stable
minority/majority ratio are possible, and (ii) a \emph{single-attractor}
regime, where only one stable minority/majority ratio exists.  In the
considered case, the crossover between these two regimes can be obtained
by an increase/decrease of the ratio $d=D_{+}/D_{-}$, provided a
subcritical population density and constant values of $\beta$ and
$\gamma$, \eqn{ratio} are fulfilled. The effect of an increasing
diffusion coefficient of one subpopulation -- or a more efficient
communication, respectively, can be understood similar to the two-box
case, \sect{4.2}.  There we have shown that larger values of $D$ decrease
 the attractor basins of some stable states for the single boxes, and
thus reduce the possibility of multiple stable states for the total system.

\section{Conclusions}
\label{6}

In this paper, we have investigated the coordination of decisions in a
spatially distributed agent community. The interaction between the agents
is described by a scalar, multi-component communication field that stores
the information about the agents' decisions.  This spatio-temporal field
is assumed to obey a reaction-diffusion equation. This way, memory
effects are modeled by means of a limited lifetime of the information,
whereas the exchange of information is modeled by means of a finite
diffusion coefficient. Dependent on the different information received at
her particular position, each agent makes a decision in a binary choice
problem. I.e. she chooses either $\{+\}$ or $\{-\}$ after comparing the
\emph{local information} available.

From analytical investigations of the mean-field case, we found that the
subpopulations of agents making a particular choice coexist at
different shares within the community. They can be found either as a
majority or a minority, provided some internal conditions (such as a
supercritical population density) are fulfilled. 

In this paper, we are mainly interested in how the majority and the
minority of agents emerge in a spatially heterogeneous system and how
they organize themselves in space. We observed that the formation of
minority/majority subpopulations goes along with a \emph{spatial
  separation process}, i.e. besides the existence of a global majority,
there are regions that are dominated by the minority. Hence, a
\emph{spatial coordination} of decisions among the agents occurs.

Further, we found that -- different from the mean-field case -- a large
range of possible global minority/majority relations can be observed that
refer to different spatial coordination patters.  We have investigated
analytically and by means of computer simulations, under which conditions
these multiple steady states occur and stable exist.  The results can be
concluded by looking at the two influencial parameters of the model,
$\kappa$ and $D$.

$\kappa$ includes the specific internal conditions within the agent
community, namely the population density $\bar{n}$, the production rate
of information per agent $s$, the lifetime of information $k$ and the
randomness $T$ that can be envisioned as a measure of the incompleteness
or incorrect transformation of information.  Defining $\nu=s\bar{n}/k$ as
the \emph{net information density}, $\kappa=2\nu/T$ describes the relation
between the mean information $\nu$ available at any
location and its efficiency $\sim 1/T$ -- in other words, the
\emph{impact} of the information produced.  We recall that the limit
\mbox{$T \to 0$} means a large impact of the available information
leading to ``rational'' decisions, whereas in the limit \mbox{$T \to
  \infty$} the influence of the information is attenuated, leading to
``random'' decisions.

In order to gain at least some impact of the available information, a
supercritical value of $\kappa>\kappa^{c}=2$ is needed. In this case the
emergence of a minority and a majority within the agent community can be
observed, whereas for $\kappa<\kappa^{c}$ only random decisions occur. To
allow \emph{multiple steady states} in the spatially extended system
instead of just one fixed minority/majority relation, $\kappa$ has to be
$\kappa^{c}<\kappa^{c}_{2}<\kappa$, where $\kappa_{2}^{c}$ itself is
increasing with the diffusion constant $D$.
Then a variety of possible spatial decision patterns can be
found, and the outcome of the decision process becomes certainly
unpredictable, \emph{both} with respect to the share of the majority
\emph{and} to the spatial distribution. 

This means that the spatial couplings, expressed in terms of $D$, are not
large enough to \emph{globally} organize the system.  Since $\kappa$
characterizes the average \emph{local} situation in a spatially extended
system in terms of a net information density, this can be also
interpreted in a way that the impact resulting from the information
\emph{exchange} does not overcome the impact resulting from the
\emph{local} information production.

For values of $\kappa$ below $\kappa_{2}^{c}(D)$, however, these local
effects become smaller, and the spatial couplings are able to organize
the whole system. Thus only one minority/majority relation occurs on the
global level, which relates to randomly different, but very similar
spatial patterns.

If we put these results in the context of a social system, we could
conclude that strong local influences, expressed in a high information
impact, can prevent the global system from being equalized and
``globalized'' by some ruling information.  While such a \emph{diversity}
might be among the wanted effects, we note again that this on the other
hand makes the system difficult to predict.

Eventually, we have addressed in this paper the influence of information
exchange on different time scales. In particular, we have assumed that
one of the two subpopulations communicates faster -- or more efficient --
than the other one, and have investigated how this affects the global
outcome of the decision process. Dependent on the ratio $d=D_{+}/D_{-}$
of the two diffusion constants we found (i) a tendency that the
subpopulation with the faster communication more likely becomes the
majority, and (ii) that the possibility of multiple steady states tends
to vanish with an increasing/decreasing $d$. In conclusion, ``efficient''
information exchange provides a suitable way to stabilize the majority
status of a particular subpopulation -- or to avoid ``diversity'' and
uncertainty in the decision process.

Finally, we want to add that the toy model of communicating agents
investigated in this paper may be easily modified or extended to describe
other processes. Without giving up the whole framework, we may consider
e.g. other types of information dissemination in the system, i.e.
\eqn{hrt} for the communication field may be replaced -- for example by a
more network-type communication among the agents.  Another possible
modification is regarding the decision process described in this paper by
means of \eqn{wh}. Here, we may envision various 
dependences on the information received from likeminded or opponent
agents.


\bibliographystyle{plain-ego-y}
\bibliography{abents-subm}


\end{document}